\documentclass[useAMS,usenatbib]{mn2e}

\usepackage{amsmath,graphicx}

\newcommand{\aap}{A\&A}
\newcommand{\apj}{ApJ}
\newcommand{\apjl}{ApJ}
\newcommand{\apjs}{ApJS}

\newcommand{\mnras}{MNRAS}
\newcommand{\na}{New Astron.}
\newcommand{\nat}{Nature}

\newcommand{\jcp}{J. Chem. Phys.}
\newcommand{\pasj}{PASJ}

\title[Radiation hydrodynamics simulations]{Multifrequency radiation hydrodynamics simulations of H$_2$ line emission in primordial, star-forming clouds}

\author[Greif]{\parbox{17.5cm}{Thomas H. Greif\thanks{E-mail: tgreif@cfa.harvard.edu}}
\\Harvard-Smithsonian Center for Astrophysics, 60 Garden Street, Cambridge, MA 02138, USA}

\begin{document}

\maketitle
\topmargin-1cm

\begin{abstract}
We investigate the collapse of primordial gas in a minihalo with three-dimensional radiation hydrodynamics simulations that accurately model the transfer of H$_2$ line emission. For this purpose, we have implemented a multiline, multifrequency ray-tracing scheme in the moving-mesh code {\sc arepo} that is capable of adaptively refining rays based on the {\sc healpix} algorithm, as well as a hybrid equilibrium/non-equilibrium primordial chemistry solver. We find that a multifrequency treatment of the individual H$_2$ lines is essential, since for high optical depths the smaller cross-section in the wings of the lines greatly increases the amount of energy that can escape. The influence of Doppler shifts due to bulk velocities is comparatively small, since systematic velocity differences in the cloud are typically smaller than the sound speed. During the initial collapse phase, the radially averaged escape fraction agrees relatively well with the fit of Ripamonti \& Abel. However, in general it is not advisable to use a simple density-dependent fitting function, since the escape fraction depends on many factors and does not capture the suppression of density perturbations due to the diffusion of radiation. The Sobolev method overestimates the escape fraction by more than an order of magnitude, since the properties of the gas change on scales smaller than the Sobolev length.
\end{abstract}

\begin{keywords}
hydrodynamics -- radiative transfer -- stars: Population~III -- galaxies: high-redshift -- cosmology: theory -- early Universe.
\end{keywords}

\section{Introduction}

In the standard $\Lambda$ cold dark matter ($\Lambda$CDM) cosmology, the first stars are expected to form at redshifts $z\ga 20$ in dark matter (DM) `minihaloes' with masses $\sim 10^6\,{\rm M_\odot}$ \citep[for recent reviews, see][]{bromm13, glover13}. The primordial gas synthesized in the big bang accretes on to these haloes and heats to the virial temperature of $T_{\rm vir}\simeq 1000\,{\rm K}$. At the centre of the halo, molecular hydrogen (H$_2$) forms via associative detachment of neutral hydrogen with negatively charged hydrogen (H$^-$), which in turn forms via radiative association of neutral hydrogen with free electrons left over after recombination \citep{sz67}. The internal ro-vibrational transitions of H$_2$ are excited by collisions with other species, and their decay produce cooling radiation that facilitates the further collapse of the gas. Detailed three-dimensional simulations of primordial star formation have shown that the onset of H$_2$ cooling in sufficiently massive haloes results in runaway collapse to a density of $n_{\rm H}\simeq n_{\rm cr}=10^4\,{\rm cm}^{-3}$, where $n_{\rm H}$ denotes the volumetric number density of hydrogen nuclei, and $n_{\rm cr}$ is the critical density at which the ro-vibrational levels of H$_2$ become populated according to local thermal equilibrium \citep[LTE;][]{abel98, bcl99, bcl02, abn00, abn02}. The gas cools to a minimum temperature of $\simeq 200\,{\rm K}$, which is set by the change in the scaling of the cooling rate from $\Lambda\propto n_{\rm H}^2$ to $\Lambda\propto n_{\rm H}$ for $n_{\rm H}\ga n_{\rm cr}$, and the microphysics of the H$_2$ molecule. The collapse rate decreases and the cloud `loiters' at this density and temperature until it has accreted a Jeans mass of $\ga 100\,{\rm M_\odot}$. The cloud then decouples from the DM potential and begins to collapse under its own gravity. At densities $n_{\rm H}\ga 10^8\,{\rm cm}^{-3}$, the H$_2$ fraction rapidly increases due to three-body reactions \citep{pss83}, and the rapidly increasing cooling rate may trigger a chemothermal instability that results in subfragmentation of the cloud \citep{sy77, silk83, ra04, yoshida06b, tao09, gsb13}. At densities $n_{\rm H}\ga 10^{10}\,{\rm cm}^{-3}$, the stability of the cloud is restored due to the increasing optical depth of the gas to H$_2$ line emission \citep{on98, ripamonti02, ra04, yoshida06b}. The second radiative coolant that becomes important is collision-induced emission, which operates at densities $n_{\rm H}\ga 10^{14}\,{\rm cm}^{-3}$ \citep{on98, ra04}. Finally, chemical cooling due to the dissociation of H$_2$ molecules precedes the near-adiabatic evolution of the gas at the highest densities \citep{on98, yoh08}.

One of the most important transitions that occurs during the initial collapse phase is the transition from optically thin to optically thick H$_2$ line cooling. Previous studies have shown that the chemical and thermal evolution of the gas at high densities depend sensitively on how rapid this transition occurs \citep{turk11, hy13}. However, an accurate solution requires the hydrodynamic evolution of the gas to be solved alongside the multifrequency radiative transfer of H$_2$ line emission. This has so far only been possible in one-dimensional calculations \citep{on98, ripamonti02}. In three-dimensional simulations, the computational cost associated with the integration of the six-dimensional photon distribution function is prohibitively expensive. Previous studies have therefore resorted to the escape probability formalism, where the optically thin cooling rate is multiplied by an escape fraction that models the probability of a photon to escape from the cloud. The escape fraction is usually derived from local properties of the gas. In \citet{ra04}, a density-dependent fit for the escape fraction was obtained from the detailed one-dimensional calculations of \citet{ripamonti02}. This method was used in \citet{tao09, tna10, turk12}. Other studies assumed that the velocity gradient in the central, Jeans-unstable cloud allows the radiation to escape relatively easily, and used the \citet{sobolev60} method to obtain the escape fraction \citep[e.g.][]{yoshida06b, clark11a, gsb13}. The Sobolev method has been traditionally applied to stellar atmospheres and molecular clouds \citep{castor70, gk74}, and is valid if the scale on which the velocity varies is much smaller than the scale on which other properties of the gas vary. In particular, it is not well suited to treat turbulent gas clouds \citep{schoenberg85, ossenkopf97}. Nevertheless, the results obtained with the Sobolev method in three-dimensional simulations agree relatively well with those of the fitting function of \citet{ra04}, even though is not clear how accurate both methods really are \citep{turk11, hy13}.

We here address this issue by performing the first three-dimensional simulations of primordial star formation that include multifrequency radiative transfer for optically thick H$_2$ line emission. In Section~\ref{sec_chem}, we present a new primordial chemistry and cooling network implemented in {\sc arepo}, and in Section~\ref{sec_rad} describe the ray-tracing scheme used to compute the radiative transfer. In Section~\ref{sec_sim}, we describe the set-up of the simulations, and in Section~\ref{sec_res} present our results. Finally, in Section~\ref{sec_sum} we summarize and draw conclusions. All distances are quoted in proper units, unless noted otherwise.

\section{Chemical solver} \label{sec_chem}

One of the most important ingredients of primordial star formation simulations is a comprehensive chemistry and cooling network. We here describe a new solver implemented in {\sc arepo} that combines a non-equilibrium solver for low densities with an equilibrium solver for high densities. This tiered approach allows us to seamlessly model the extremely large dynamic range of more than $20$ orders of magnitude in density that builds up in self-gravitating, primordial gas clouds.

\subsection{Methodology}

We use an operator-split approach to solve for the coupled evolution of the chemical abundances and internal energy of the gas, represented by a system of first-order differential equations:
\begin{equation}
{\dot {\bf y}}={\bf F}(t, {\bf y}),
\end{equation}
where ${\bf y}$ denotes the chemical species and internal energy, ${\bf F}$ is a function that incorporates the rate equations, and $t$ denotes the time. The time dependency implicitly includes all external dependencies, such as on density and redshift. For a given hydrodynamic time step $\Delta t$, the above differential equation is integrated using the publicly available solver {\sc sundials cvode}, which employs a variable order, variable step multistep backward differencing scheme \citep{hindmarsh05}. We model three independent chemical species: H$^-$, H$_2$, and H$^+$. The abundance of H$^-$ may be trivially derived from that of H$^+$, such that only the abundances of H$_2$ and H$^+$ are included in ${\bf y}$. The electron abundance is trivially given by $y_{\rm e}=y_{\rm H^+}$, and that of neutral hydrogen by $y_{\rm H\,{\textsc i}}=1-2y_{\rm H_2}-y_{\rm H^+}$, where $y_X=n_X/n_{\rm H}$ denotes the ratio of the number density of chemical species $X$ to the number density of hydrogen nuclei. The latter is given by $n_{\rm H}=X_{\rm H}\rho/m_{\rm H}$, where $X_{\rm H}=0.76$ is the cosmological mass fraction of hydrogen, $\rho$ the volumetric mass density, and $m_{\rm H}$ the mass of the hydrogen atom. Since reactions involving helium are comparatively unimportant at the densities and temperatures modelled here, we assume that it remains chemically inert and is in the ground electronic state. Because of the tight coupling of some of the chemical rates to the internal energy, a relative accuracy of $f_{\rm acc,neq}=10^{-4}$ is necessary to avoid significant spurious oscillations. Below $y_{\rm abs}=10^{-20}$, the chemical species are no longer evolved accurately, which avoids an unnecessary computational overhead.

\begin{table*}
\centering
\caption{Chemistry and cooling network, and the corresponding rate coefficients. $T_{\rm K}$ denotes the temperature in units of K, $T_{\rm eV}$ the temperature in units of eV, and $T_{\rm CMB,K}=2.725(1+z)^4$ the temperature of the CMB in units of K. The equilibrium constants $q_{\rm H_2}$ and $q_{\rm H}$ are introduced in Section~\ref{sec_rates}.}
\begin{tabular}{lllll} \hline \hline
No. & Process & Rate coefficient & Units & Reference \\ \hline
1 & ${\rm H}+{\rm e}^-\rightarrow{\rm H^-}+\gamma$ & $1.4\times 10^{-18}T_{\rm K}^{0.928}\exp{(-T_{\rm K}/1.62\times 10^4)}$ & cm$^3$\,s$^{-1}$ & \citet{gp98} \\
2 & ${\rm H}+{\rm H}^-\rightarrow{\rm H_2}+{\rm e}^-$ & $1.35\times 10^{-9}(T_{\rm K}^{0.098493}+0.32852T_{\rm K}^{0.5561}$ & cm$^3$\,s$^{-1}$ & \citet{kreckel10} \\
& & $+2.771\times 10^{-7}T_{\rm K}^{2.1826})/(1+6.191\times 10^{-3}T_{\rm K}^{1.0461}$ & \\
& & $+8.9712\times 10^{-11}T_{\rm K}^{3.0424}+3.2576\times 10^{-14}T_{\rm K}^{3.7741})$ & \\
3 & $3{\rm H}\rightarrow{\rm H}+{\rm H_2}$ & $6\times 10^{-32}T_{\rm K}^{-0.25}+2\times 10^{-31}T_{\rm K}^{-0.5}$ & cm$^6$\,s$^{-1}$ & \citet{forrey13} \\
4 & $2{\rm H}+{\rm H_2}\rightarrow 2{\rm H_2}$ & $k_3/8$ & cm$^6$\,s$^{-1}$ & \citet{jgc67} \\
5 & ${\rm H}+{\rm H_2}\rightarrow 3{\rm H}$ & $q_{\rm H_2}k_3$ & cm$^3$\,s$^{-1}$ & Detailed balance with $(3)$\\
6 & $2{\rm H_2}\rightarrow 2{\rm H}+{\rm H_2}$ & $q_{\rm H_2}k_4$ & cm$^3$\,s$^{-1}$ & Detailed balance with $(4)$\\
7 & ${\rm H_2}+\gamma\rightarrow 2{\rm H}$ & $1.38\times 10^{-12}$ & s$^{-1}$ & \citet{abel97} \\
8 & ${\rm H}+{\rm e}^-\rightarrow{\rm H^+}+2{\rm e}^-$ & ${\rm exp}[-32.71396786+13.536556\ln{T_{\rm eV}}$ & cm$^3$\,s$^{-1}$ & \citet{jle87}\\
& & $-5.73932875(\ln{T_{\rm eV}})^2+1.56315498(\ln{T_{\rm eV}})^3$ & \\
& & $-0.2877056(\ln{T_{\rm eV}})^4+0.0348255977(\ln{T_{\rm eV}})^5$ & \\
& & $-2.63197617\times 10^{-3}(\ln{T_{\rm eV}})^6$ & \\
& & $+1.11954395\times 10^{-4}(\ln{T_{\rm eV}})^7$ & \\
& & $-2.03914985\times 10^{-6}(\ln{T_{\rm eV}})^8]$ & \\
9 & ${\rm H^+}+{\rm e}^-\rightarrow{\rm H}+\gamma$ & $2.753\times 10^{-14}(3.15614\times 10^5/T_{\rm K})^{1.5}$ & cm$^3$\,s$^{-1}$ & \citet{ferland92}\\
& & $\times(1+(1.15188\times 10^5/T_{\rm K})^{0.407})^{-2.242}$ & \\
10 & ${\rm H^+}+{\rm e}^-\rightarrow{\rm H}+\gamma$ & $k_8/q_{\rm H}$ & cm$^6$\,s$^{-1}$ & Detailed balance with (8) \\
11 & H$_2$ line cooling & ${\rm dex}[-103+97.59{\rm log}T_{\rm K}-48.05({\rm log}T_{\rm K})^2$ & erg\,cm$^3$\,s$^{-1}$ & \citet{gp98} \\
& & $+10.8({\rm log}T_{\rm K})^3-0.9032({\rm log}T_{\rm K})^4]$ & \\
12 & H$_2$ CIE & $5.3\times 10^{-49}T_{\rm K}^4$ & erg\,cm$^3$\,s$^{-1}$ & \citet{ra04} \\
13 & Ly$\alpha$ cooling & $7.5\times 10^{-19}\exp{(-1.18348\times 10^5/T_{\rm K})}$ & erg\,cm$^3$\,s$^{-1}$ & \citet{cen92} \\
& & $/\left[1+(T_{\rm K}/10^5)^{0.5}\right]$ & \\
14 & IC cooling & $5.65\times 10^{-36}T_{\rm CMB,K}^4(T_{\rm K}-T_{\rm CMB,K})$ & erg\,s$^{-1}$ & \citet{peebles71}
\end{tabular}
\end{table*}

\subsection{Chemistry} \label{sec_rates}

The chemical and thermal processes included in our reaction network are shown in Table~1, together with the corresponding rate coefficients and references. To enable fast look-up, the rates are linearly interpolated from a table with $5000$ logarithmically spaced temperature bins between $10$ and $10^8\,{\rm K}$. Because of the comparatively large reaction rates, we assume that H$^-$ is in chemical equilibrium:
\begin{equation}
y_{\rm H^-}=\frac{k_1}{k_2}y_{\rm H^+},
\end{equation}
where $k_1$ denotes the rate coefficient for the formation of H$^-$ via radiative association of H\,{\sc i} and free electrons, and $k_2$ the rate coefficient for the destruction of H$^-$ by associative detachment with H\,{\sc i}. The latter reaction results in the formation of H$_2$. Molecular hydrogen may also be formed by three-body reactions involving three hydrogen atoms or two hydrogen atoms and one hydrogen molecule, while it is destroyed by collisions with hydrogen atoms and molecules, as well as radiation in the Lyman--Werner (LW) bands:
\begin{align} \label{eq_h2}
{\dot y}_{\rm H_2}= & \;k_2 y_{\rm H^-}y_{\rm H\,{\textsc i}}n_{\rm H}+(k_3 y^3_{\rm H\,{\textsc i}}+k_4 y_{\rm H_2}y_{\rm H\,{\textsc i}}^2)n_{\rm H}^2 \notag \\
& \;-k_5 y_{\rm H_2}y_{\rm H\,{\textsc i}}n_{\rm H}-k_6 y^2_{\rm H_2}n_{\rm H}-k_7J_{21}f_{\rm sh}y_{\rm H_2},
\end{align}
where $J_{21}$ denotes the specific intensity in the LW bands in units of $10^{-21}\,{\rm erg}\,{\rm s}^{-1}\,{\rm cm}^{-2}\,{\rm Hz}^{-1}\,{\rm sr}^{-1}$. We use the recently updated rate coefficients for associative detachment and three-body H$_2$ formation of \citet{kreckel10} and \citet{forrey13}, respectively. The shielding factor for incident LW radiation is given by
\begin{align}
f_{\rm sh}= & \;\frac{0.965}{(1+x/b_5)^{1.1}}+\frac{0.035}{(1+x)^{0.5}} \\
& \;\times\exp{[-8.5\times 10^{-4}(1+x)^{0.5}]},\notag
\end{align}
where $x=N_{\rm H_2,eff}/5\times 10^{14}\,{\rm cm}^{-2}$, $N_{\rm H_2,eff}$ is the effective H$_2$ column density (see Section~\ref{sec_sob}), $b_5=v_{\rm th}/\left({\rm km}\,{\rm s}^{-1}\right)$, and $v_{\rm th}$ the thermal velocity of the H$_2$ molecules:
\begin{equation}
v_{\rm th}=\sqrt{2k_{\rm B}T/m_{\rm H_2}}.
\end{equation}
Here, $m_{\rm H_2}$ denotes the mass of the H$_2$ molecule, $k_{\rm B}$ Boltzmann's constant, and $T$ the temperature of the gas \citep{db96, whb11}.

The collisional dissociation rates are obtained from the three-body formation rates by applying the principle of detailed balance:
\begin{equation}
k_{\rm diss}=q_{\rm H_2}k_{\rm 3b},
\end{equation}
where $q_{\rm H_2}$ is the equilibrium constant:
\begin{equation} \label{eq_eq_h2}
q_{\rm H_2}=\frac{n_{\rm H\,{\textsc i}}^2}{n_{\rm H_2}}=\frac{Z_{\rm H}^2}{Z_{\rm H_2}}\left(\frac{\pi m_{\rm H}k_{\rm B}T}{h_{\rm p}^2}\right)^{3/2}\,\exp{(-\chi_{\rm H_2}/k_{\rm B}T)},
\end{equation}
and $h_{\rm p}$ denotes Planck's constant, $\chi_{\rm H_2}=4.48\,{\rm eV}$ the binding energy of H$_2$, and $Z_{\rm H}$ and $Z_{\rm H_2}$ the partition functions of atomic and molecular hydrogen, respectively:
\begin{equation}
Z=\sum_{i}g_i\exp{\left(-\frac{E_i}{k_{\rm B}T}\right)},
\end{equation}
where $g_i$ is the degeneracy of state $i$ with energy $E_i$. In the case of atomic hydrogen, $g_i=2i^2$, $E_i=\chi_{\rm H}/i^2$, and $\chi_{\rm H}=13.6\,{\rm eV}$ denote the ionization energy of atomic hydrogen, and the sum is truncated at $i=5$. In the case of H$_2$, only the ro-vibrational transitions of the electronic ground state are modelled, such that $g_i=2J_i+1$, where $J$ denotes the rotational quantum number. The energy levels are taken from \citet{bfm89}, using vibrational quantum numbers $0\leq v\leq 2$ and rotational quantum numbers $0\leq J< 20$, which is sufficient for the regime in which H$_2$ cooling is important.

The formation and destruction of ionized hydrogen is governed by collisional ionization with electrons and the inverse process, recombinations. The recombination rate at high densities is obtained by applying the principle of detailed balance to the collisional ionization rate of \citet{jle87} :
\begin{equation}
k_{\rm rec}=k_{\rm coll}/q_{\rm H},
\end{equation}
where
\begin{equation}
q_{\rm H}=\frac{n_{\rm H^+}^2}{n_{\rm H\,{\textsc i}}}=\frac{2}{Z_{\rm H}}\left(\frac{2\pi m_{\rm e} k_{\rm B}T}{h^2_{\rm p}}\right)^{1.5}\,\exp{(-\chi_{\rm H}/k_{\rm B}T)},
\end{equation}
and $m_{\rm e}$ is the mass of the electron. At low densities, the case B recombination rate of \citet{ferland92} is used, which includes recombinations to all levels except the ground state. The two regimes are smoothly adjoined by a transition function, such that the net H$^+$ formation rate is given by
\begin{equation}
{\dot y}_{\rm H^+}=k_8 y_{\rm H^+}y_{\rm H\,{\textsc i}}n_{\rm H}-(k_9 y^2_{\rm H^+}n_{\rm H})^d (k_{10} y^3_{\rm H^+} n_{\rm H}^2)^{1-d},
\end{equation}
where
\begin{equation}
d=\frac{1}{1+n_{\rm H}/n_{\rm H, trans}},
\end{equation}
and $n_{\rm H, trans}=10^{17}\,{\rm cm}^{-3}$. This ensures that the H$^+$ abundance approaches the thermal equilibrium abundance at high densities, since we do not include the inverse reactions for recombinations to all hydrogen levels (reaction~9 in Table~1).

\subsection{Heating and cooling}

The rate of change of the volumetric internal energy density $u$ is given by
\begin{equation}
{\dot u}=\Gamma_{\rm chem}-\Lambda_{\rm cool},
\end{equation}
where $\Gamma_{\rm chem}$ denotes heating due to chemical processes, and $\Lambda_{\rm cool}$ cooling due to radiative processes. We only include the chemical heating and cooling of the gas due to the formation and destruction of H$_2$, since significant amounts of H$^+$ are formed only at very high densities, where the abundances become inaccurate due to various non-ideal gas effects \citep{ripamonti02}. The chemical heating rate may therefore be conveniently written as
\begin{equation}
\Gamma_{\rm chem}=\chi_{\rm H_2}{\dot y}_{\rm H_2}n_{\rm H}.
\end{equation}
The radiative cooling rate is given by
\begin{equation}
\Lambda_{\rm cool}=\Lambda_{\rm line}+\Lambda_{\rm CIE}+\Lambda_{\rm Ly\alpha}+\Lambda_{\rm IC},
\end{equation}
and includes H$_2$ line cooling, H$_2$ collision-induced emission, Ly$\alpha$ cooling, and inverse Compton scattering of electrons with cosmic microwave background (CMB) photons. The H$_2$ line cooling rate is discussed in detail in Section~\ref{sec_line}. The cooling rate due to collision-induced emission is given by
\begin{equation}
\Lambda_{\rm CIE}=k_{12}y_{\rm H_2}n_{\rm H}^2f_{\rm esc},
\end{equation}
where $k_{12}$ and the escape fraction are taken from \citet{ra04}. The latter is given by
\begin{equation}
f_{\rm esc}=\frac{1-\exp{(-\tau_{\rm cont})}}{\tau_{\rm cont}},
\end{equation}
where
\begin{equation}
\tau_{\rm cont}=\left(\frac{n_{\rm H}}{n_{\rm H,cont}}\right)^{2.8}
\end{equation}
and $n_{\rm H,cont}=1.5\times 10^{16}\,{\rm cm}^{-3}$. The Ly$\alpha$ cooling rate is given by:
\begin{equation}
\Lambda_{\rm Ly\alpha}=k_{13}y_{\rm H^+}y_{\rm H\,{\textsc i}}n_{\rm H}^2f_{\rm esc},
\end{equation}
where $k_{13}$ is taken from \citet{cen92}, and
\begin{equation}
f_{\rm esc}=\exp{(-n_{\rm H}/n_{\rm H,cont})}.
\end{equation}
The latter approximately reproduces the temperature--density relation found in the one-zone models of \citet{omukai01}. Finally, the IC cooling rate is given by
\begin{equation}
\Lambda_{\rm IC}=k_{14}y_{\rm H^+}n_{\rm H},
\end{equation}
where $k_{14}$ is taken from \citet{peebles71}.

\subsection{H$_2$ line cooling} \label{sec_line}

The H$_2$ line cooling rate is obtained by adjoining the rate in the limit $n_{\rm H}\rightarrow 0$ with the LTE rate \citep{gp98}:
\begin{equation}
\Lambda_{\rm line}=\frac{\Lambda_{\rm LTE}}{1+\Lambda_{\rm LTE}/\Lambda_{n\rightarrow 0}}.
\end{equation}
In the above equation, the low-density rate is given by
\begin{equation}
\Lambda_{\rm n\rightarrow 0}=k_{11}y_{\rm H\,{\textsc i}}y_{\rm H_2}n^2_{\rm H},
\end{equation}
where $k_{11}$ is taken from \citet{gp98}. If the gas is optically thin, the cooling rate may be calculated directly from the Einstein coefficients corresponding to the individual ro-vibrational transitions of H$_2$:
\begin{equation}
\Lambda_{\rm LTE, thin}=\sum_{\rm u,l}E_{\rm ul}A_{\rm ul}n_{\rm u}
\end{equation}
where $E_{\rm ul}$ denotes the energy emitted by the transition from the upper state u to the lower state l, $A_{\rm ul}$ the Einstein coefficient, and $n_{\rm u}$ the number density of H$_2$ molecules in the upper state. The Einstein coefficients are taken from \citet{tkd77}. The relative numbers of H$_2$ molecules in the upper and lower states are given by
\begin{equation}
B_{\rm u,l}=\frac{n_{\rm u,l}}{n_{\rm H_2}}=\frac{g_{\rm u,l}}{Z_{\rm H_2}}\exp{\left(-\frac{E_{\rm u,l}}{k_{\rm B}T}\right)}.
\end{equation}
This allows the cooling rate per H$_2$ molecule to be tabulated as a function of temperature:
\begin{equation}
\epsilon_{\rm LTE,thin}=\sum_{\rm u,l}E_{\rm ul}A_{\rm ul}B_{\rm u}.
\end{equation}

\subsubsection{Fitting function} \label{sec_fit}

\citet{ra04} assumed that the optically thick cooling rate can be obtained from the optically thin cooling rate via
\begin{equation}
\epsilon_{\rm LTE}=f_{\rm esc}\epsilon_{\rm LTE, thin},
\end{equation}
where
\begin{equation} \label{eq_fit1}
f_{\rm esc}=\frac{bx}{x^{b}+b-1}
\end{equation}
for $x\ge 1$ and $f_{\rm esc}=1$ for $x<1$. The parameter $x$ is given by
\begin{equation}
x=b^{1/\left(b-1\right)}\frac{n_{\rm H}}{X_{\rm H}n_{\rm H,line}},
\end{equation}
where $b=1.45$ and $n_{\rm H,line}=8\times 10^9\,{\rm cm}^{-3}$. This formula reproduces the slope of the fit to the detailed one-dimensional calculations of \citet{ripamonti02}, but has the advantage of a continuous derivative at $x=1$.

\subsubsection{Sobolev method} \label{sec_sob}

\citet{yoshida06b} were the first to use the Sobolev method to model optically thick H$_2$ line transfer in primordial gas clouds. They computed the escape fraction for each individual transition:
\begin{equation}
\epsilon_{\rm LTE}=\sum_{\rm u,l}E_{\rm ul}A_{\rm ul}B_{\rm u}f_{\rm esc,ul},
\end{equation}
where
\begin{equation} \label{eq_tau}
f_{\rm esc,ul}=\frac{1-\exp{\left(-\tau_{\rm ul}\right)}}{\tau_{\rm ul}},
\end{equation}
and
\begin{equation} \label{eq_tau_ul}
\tau_{\rm ul}=\sigma_{\rm ul}n_{\rm l} L_{\rm Sob}.
\end{equation}
Here, $\sigma_{\rm ul}$ denotes the cross-section for the transition ${\rm u}\rightarrow {\rm l}$, and $L_{\rm Sob}$ the Sobolev length. The cross-section corrected for stimulated emission is given by
\begin{equation} \label{eq_cross}
\sigma_{\rm ul}=\frac{1}{8\pi}\left(\frac{h_{\rm p}c}{E_{\rm ul}}\right)^2\frac{g_{\rm u}}{g_{\rm l}}A_{\rm ul}\left[1-\exp{\left(-\frac{E_{\rm ul}}{k_{\rm B}T}\right)}\right]\phi_{\rm ul},
\end{equation}
where $c$ denotes the speed of light, and $\phi_{\rm ul}$ is the line profile for thermal Doppler broadening:
\begin{equation}
\phi_{\rm ul}=\frac{1}{\sqrt{\pi}\Delta\nu_{\rm D,ul}}\exp{\left[-\left(\frac{\nu -\nu_{\rm ul}}{\Delta\nu_{\rm D,ul}}\right)^2\right]}.
\end{equation}
In the above equation, $\nu$ denotes the frequency, $\nu_{\rm ul}=E_{\rm ul}/h_{\rm p}$, and $\Delta\nu_{\rm D,ul}$ is the thermal Doppler broadening parameter:
\begin{equation} \label{eq_nud}
\Delta\nu_{\rm D,ul}=\nu_{\rm ul}\frac{v_{\rm th}}{c}.
\end{equation}
Unless noted otherwise, terms involving the cross-section are evaluated at the centre of the line. \citet{yoshida06b} computed the Sobolev length and escape fraction along the principal axes of the computational domain, and averaged the escape fractions along the different directions to obtain the overall escape fraction. We here follow \citet{clark11a} and compute the Sobolev length as
\begin{equation}
L_{\rm Sob}={\rm min}\left(\frac{v_{\rm th}}{\left|{\nabla}{\bf v}\right|},L_{\rm Jeans}\right),
\end{equation}
where ${\bf v}$ is the velocity of the gas, and $L_{\rm Jeans}$ the Jeans length. The limiter in the above equation ensures that the Sobolev length remains smaller than the Jeans length if the divergence of the velocity is small. To speed up the computation of the optically thick cooling rate, we follow \citet{clark11a} and write equation~\ref{eq_tau_ul} as
\begin{equation}
\tau_{\rm ul}=\sigma_{\rm ul}B_{\rm l}N_{\rm H_2, eff},
\end{equation}
where $N_{\rm H_2, eff}$ is the effective H$_2$ column density:
\begin{equation}
N_{\rm H_2, eff}=n_{\rm H_2}L_{\rm Sob}.
\end{equation}
The optically thick cooling rate can thus be tabulated as a function of temperature and column density. For the latter, we use $200$ logarithmically spaced bins between $10^{21}$ and $10^{30}\,{\rm cm}^{-2}$.

\subsubsection{Ray tracing} \label{sec_ray}

Finally, the frequency-dependent radiative transfer of H$_2$ line emission may be computed accurately and self-consistently with the ray-tracing method described in Section~\ref{sec_rad}. For computational efficiency, the cooling rate per H$_2$ molecule due to each transition is tabulated as a function of temperature:
\begin{equation} \label{eq_eps_h2}
\epsilon_{\rm LTE,ul}=E_{\rm ul}A_{\rm ul}B_{\rm u}.
\end{equation}
We also tabulate the effective cross-section:
\begin{equation} \label{eq_cross_eff}
\sigma_{\rm eff,ul}=\sigma_{\rm ul}B_{\rm l}
\end{equation}
and the level-averaged cross-section:
\begin{equation} \label{eq_cross_avg}
\sigma_{\rm eff}=\frac{\sum_{\rm u,l}\epsilon_{\rm LTE,ul}\sigma_{\rm eff,ul}}{\sum_{\rm u,l}\epsilon_{\rm LTE,ul}}.
\end{equation}
These are used to compute the attenuation of the radiation.

\subsection{Adiabatic index}

The pressure, temperature, and internal energy density of the gas are related via
\begin{equation}
P=\rho\frac{k_{\rm B}T}{\mu m_{\rm H}}=(\gamma -1)u,
\end{equation}
where $\mu$ and $\gamma$ denote the mean molecular weight and adiabatic index of the gas, respectively. The latter is given by
\begin{equation}
\frac{1}{\gamma -1}=\frac{\sum_i y_i/(\gamma_i -1)}{\sum_i y_i},
\end{equation}
where the sum extends over all chemical species. In our case, the adiabatic index is given by
\begin{equation}
\frac{1}{\gamma -1}=\frac{1+y_{\rm He}-2y_{\rm H_2}+y_{\rm H^+}}{y_n\left(\gamma_{\rm m}-1\right)}+\frac{y_{\rm H_2}}{y_n\left(\gamma_{\rm H_2}-1\right)},
\end{equation}
where $y_n=1+y_{\rm He}-y_{\rm H_2}+y_{\rm H^+}$ and $y_{\rm He}=\left(1/X_{\rm H}-1\right)/4$. The adiabatic index for a monatomic gas is given by $\gamma_{\rm m}=5/3$, and for H$_2$ by
\begin{equation}
\frac{1}{\gamma_{\rm H_2}-1}=\frac{5}{2}+\frac{x^2{\rm e}^x}{\left({\rm e}^x-1\right)^2},
\end{equation}
where $x=6.1\times 10^3\,{\rm K}/T$. The second term in this equation accounts for the vibrational degrees of freedom of H$_2$ \citep{yoshida06b}. In analogy to the adiabatic index, the mean molecular weight of the gas is given by
\begin{equation}
\mu=\frac{\sum_i m_i y_i}{m_{\rm H}\sum_i y_i},
\end{equation}
where $m_i$ denotes the particle mass of species $i$. Here, this simplifies to
\begin{equation}
\mu =\frac{1+4y_{\rm He}}{1+y_{\rm He}-y_{\rm H_2}+y_{\rm H^+}}.
\end{equation}

\subsection{Equilibrium chemistry}

Once the density exceeds $n_{\rm H_2,eq}=10^{15}\,{\rm cm}^{-3}$, the H$_2$ abundance may be safely assumed to be in thermal equilibrium \citep{on98}. This simplifies the chemistry, since updating the H$_2$ abundance only requires the solution of an implicit equation instead of a coupled differential equation. For a new internal energy, the updated H$_2$ abundance must be consistent with the chemical heating and cooling of the gas due to the formation and dissociation of H$_2$:
\begin{equation}
u-u_{\rm init}=\chi_{\rm H_2}(y_{\rm H_2}-y_{\rm H_2,init})n_{\rm H},
\end{equation}
where $u_{\rm init}$ and $y_{\rm H_2, init}$ are the internal energy density and H$_2$ abundance at the beginning of the time step. According to equation~\ref{eq_eq_h2}, the H$_2$ abundance is related to the temperature via
\begin{equation}
\frac{\left(1-2y_{\rm H_2}-y_{\rm H^+}\right)^2}{y_{\rm H_2}}=\frac{q_{\rm H_2}}{n_{\rm H}},
\end{equation}
which is solved with a bisection method that uses $u=(u_{\rm max}-u_{\rm min})/2$ as an initial guess for the internal energy. Here, $u_{\rm min}=f_{\rm bi}u_{\rm init}$, $u_{\rm max}=u_{\rm init}/f_{\rm bi}$, and $f_{\rm bi}=0.1$ gives robust minimum and maximum values. 

For each new guess of the internal energy, the adiabatic index, mean molecular weight, and temperature are updated using the H$_2$ fraction at the beginning of the time step. This does not result in a substantial error, since the adiabatic index and mean molecular weight do not change much over a single time step. The updated temperature is used to compute the equilibrium constant $q_{\rm H_2}$, which is then used to solve the above equation. The physically meaningful solution is the negative branch of
\begin{equation}
y_{\rm H_2}=\frac{-B\pm E}{2A},
\end{equation}
where the coefficients are given by
\begin{align}
A= & \;4, \\
B= & \;-4\left(1-y_{\rm H^+}\right)-q_{\rm H_2}/n_{\rm H}, \\
C= & \;\left(1-y_{\rm H^+}\right)^2, \\
D= & \;4AC, \\
E= & \;\left(B^2-D\right)^{1/2}.
\end{align}
A subtle problem arises if $D$ is much smaller than $B^2$. In this case, $E$ may be truncated due to rounding errors, which results in $E=B^2$ and $y_{\rm H_2}=0$. This unphysical solution is avoided by a Taylor expansion of $E$ around $D=0$:
\begin{equation}
E=\left|B\right|-\frac{D}{2\left|B\right|}.
\end{equation}
For $D<10^{-10}B^2$, the H$_2$ abundance is therefore given by
\begin{equation}
y_{\rm H_2}=\frac{D}{4A\left|B\right|},
\end{equation}
where we have exploited the fact that $B<0$. Once the new H$_2$ abundance has been obtained, the new internal energy density is used as a solution if $\left(u_{\rm new}-u_{\rm init}\right)/u_{\rm init}<f_{\rm acc,eq}=10^{-7}$. This comparatively high accuracy is necessary to obtain a self-consistent solution at the very high densities and temperatures within metal-free protostars. After the equilibrium step is completed, the non-equilibrium solver is used to update the H$^+$ abundance as well as the internal energy density, which is subject to radiative cooling. The equilibrium and non-equilibrium steps are subcycled on a time step:
\begin{equation}
\Delta t_{\rm sub}=f_{\rm sub}\Delta t,
\end{equation}
where we have found that $f_{\rm sub}=0.2$ suppresses visible fluctuations.

For $n_{\rm H}\ga n_{\rm H^+,eq}=10^{18}\,{\rm cm}^{-3}$, the H$^+$ abundance also converges to the thermal equilibrium value \citep{on98}. As opposed to H$_2$, we do not account for the chemical heating and cooling of the gas due to changes in the H$^+$ abundance. A further simplification arises due to the fact that the H$_2$ abundance generally decreases to well below unity as the H$^+$ abundance increases to unity. The resulting equation is therefore comparatively simple:
\begin{equation}
\frac{y_{\rm H^+}^2}{1-y_{\rm H^+}}=\frac{q_{\rm H}}{n_{\rm H}}.
\end{equation}
The root of this equation is found using the same bisection method as for the H$_2$ abundance, but with $f_{\rm bi}=0.01$. Since the H$_2$ abundance may depend sensitively on the H$^+$ abundance, the latter is updated first.

\subsection{Test calculations}

We investigate the accuracy and reliability of the chemistry network with an idealized dynamical model for self-gravitating, primordial gas clouds. We assume that the clouds are uniform, spherically symmetric, and collapse at the free-fall rate. In this case, the density increases according to
\begin{equation}
{\dot \rho}=\rho/t_{\rm ff},
\end{equation}
where $t_{\rm ff}$ denotes the free-fall time:
\begin{equation}
t_{\rm ff}=\left(\frac{3\pi}{32G\rho}\right)^{1/2},
\end{equation}
and $G$ is the gravitational constant. The internal energy density evolves according to
\begin{equation}
{\dot u}=\gamma u/t_{\rm ff}-\Lambda,
\end{equation}
where the first term denotes the adiabatic heating rate due to the collapse of the cloud, and the second term the net cooling rate due to all other processes. The clouds are set up with an initial temperature of $200\,{\rm K}$ at $n_{\rm H}=10^{-3}\,{\rm cm}^{-3}$, a redshift of $20$, and initial abundances $y_{\rm H_2}=6.6\times 10^{-7}$ and $y_{\rm H^+}=2.6\times 10^{-4}$. These abundances are also used in the full three-dimensional simulations. The effective H$_2$ column density is obtained using the Jeans length instead of the Sobolev length.

\begin{figure}
\begin{center}
\includegraphics[width=8cm]{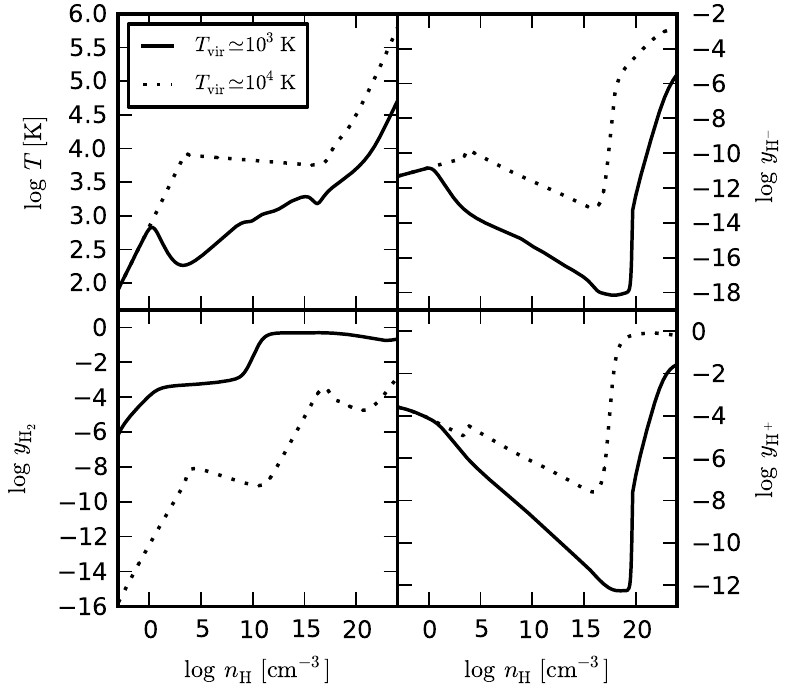}
\caption{From top left to bottom right: temperature, H$^-$ abundance, H$_2$ abundance, and H$^+$ abundance versus density in a simple one-zone dynamical collapse model. The solid line represents a minihalo, and the dotted line an atomic cooling halo, where H$_2$ cooling is suppressed by a LW background. The profiles agree well with more detailed studies \citep[e.g.][]{on98, ripamonti02}.}
\label{fig_chem}
\end{center}
\end{figure}

In Fig.~\ref{fig_chem}, we show the evolution of the cloud when H$_2$ cooling operates (solid line), and when it is suppressed up to the critical density $n_{\rm cr}=10^4\,{\rm cm}^{-3}$ by a LW background with a strength of $J_{21}=10^5$ (dotted line). The former case represents the collapse of the gas in a minihalo with $T_{\rm vir}\simeq 10^3\,{\rm K}$, and the latter in an atomic cooling halo with $T_{\rm vir}\simeq 10^4\,{\rm K}$. The resulting temperature and abundance profiles agree well with those found in previous studies \citep[e.g.][]{on98, omukai01, ripamonti02}. In the minihalo, the characteristic drop in temperature to $\simeq 200\,{\rm K}$ at $n_{\rm H}\simeq n_{\rm cr}$ is followed by a gradual increase to $\simeq 10^4\,{\rm K}$ over many orders of magnitude in density. In the atomic cooling halo, the gas remains nearly isothermal with $T\simeq 10^4\,{\rm K}$ up to a density of $n_{\rm H}\simeq 10^{16}\,{\rm cm}^{-3}$, where the gas becomes optically thick to continuum radiation.

\section{Radiative transfer} \label{sec_rad}

We here present a new multiline, multifrequency ray-tracing scheme that is capable of solving the static radiative transfer equation for point sources as well as diffuse emission. We describe the methodology and its application to H$_2$ line transfer in primordial gas clouds, and show the results of a few idealized test simulations that demonstrate the accuracy and numerical convergence of the method.

\subsection{Ray tracing}

Similar to the Euler equations for a fluid consisting of massive particles, the radiative transfer equation for an ensemble of photons may be derived rigorously from the Boltzmann equation. The photon distribution function $f_\gamma$ is a function of position ${\bf x}$ and momentum ${\bf p}$, and evolves according to
\begin{equation}
\frac{{\rm d}f_\gamma}{{\rm d}t}=\left.\frac{{\rm d}f_\gamma}{{\rm d}t}\right|_{\rm int},
\end{equation}
where the term on the right-hand side denotes the change in $f_\gamma$ due to the creation, destruction, or re-distribution (scattering) of photons by interactions with other particles. The left-hand side of this equation may be expanded to:
\begin{equation}
\frac{{\rm d}f_\gamma}{{\rm d}t}=\frac{\partial f_\gamma}{\partial t}+\frac{c}{a}{\bf{\hat n}}\nabla f_\gamma+{\bf{\dot p}}\frac{\partial f_\gamma}{\partial {\bf p}},
\end{equation}
where $a$ denotes the scale factor, and ${\bf{\hat n}}$ the normalized propagation direction of the photons. If relativistic effects are ignored, then $a=1$ and ${\bf{\dot p}}=0$, such that
\begin{equation}
\frac{\partial f_\gamma}{\partial t}+c{\bf{\hat n}}\nabla f_\gamma=\left.\frac{{\rm d}f_\gamma}{{\rm d}t}\right|_{\rm int}.
\end{equation}
This is the classical radiative transfer equation in terms of the photon distribution function. Assuming that the interaction term varies on time-scales significantly larger than the light-crossing time, the time-dependent term on the left-hand side of the equation may be omitted, which yields
\begin{equation}
{\bf{\hat n}}\nabla f_\gamma=\frac{1}{c}\left.\frac{{\rm d}f_\gamma}{{\rm d}t}\right|_{\rm int}.
\end{equation}
If the gradient of the photon distribution function is projected along the propagation direction, the well-known ray-tracing equation is obtained:
\begin{equation}
\frac{\partial f_\gamma}{\partial r}=\frac{1}{c}\left.\frac{{\rm d}f_\gamma}{{\rm d}t}\right|_{\rm int},
\end{equation}
where $r$ is the spatial coordinate along the ray. For pure absorption, this simplifies to
\begin{equation}
\frac{\partial f_\gamma}{\partial r}=-\alpha f_\gamma,
\end{equation}
where $\alpha=\sigma n$ is the absorption coefficient. This equation is trivially solved by
\begin{equation} \label{eq_fgamma}
f_\gamma=f_{\gamma,{\rm init}}\exp{\left(-\tau\right)},
\end{equation}
where $f_{\gamma,{\rm init}}$ is the initial photon distribution function, and
\begin{equation}
\tau=\int_0^r\alpha {\rm d}r'.
\end{equation}
Since photon numbers are conserved, equation~\ref{eq_fgamma} may also be written as
\begin{equation}
N_\gamma=N_{\gamma,{\rm init}}\exp{\left(-\tau\right)},
\end{equation}
where $N_\gamma$ is the number of photons in a given ray and frequency range.

\subsection{Implementation}
Similar to the cooling or heating of the gas due to chemical processes, the heating of the gas due to radiation is modelled by an additional source term in the energy equation of the Euler equations. We use an operator-split approach, such that the radiative heating rate is obtained from a ray-tracing step that succeeds the hydrodynamic step. Since a complete Voronoi tessellation of the computational domain is necessary to ensure that the neighbour lists that are used in the ray walk are complete, all cells must be evolved on the same time step. Point sources may be initialized from a pre-defined list of points or the positions of individual cells. For diffuse radiation, a fraction $f_{\rm src}$ of all cells in the computational domain become sources. For each source, $N_{\rm rays,init}=12\times 4^{l_{\rm init}}$ rays are cast using the {\sc healpix} algorithm to find their position on the unit sphere, where $l_{\rm init}$ is the initial {\sc healpix} level \citep{gorski05}. The rays from each source are then rotated using a random sample of the Euler angles. For monochromatic radiation with a frequency $\nu$, the initial photon numbers are given by
\begin{equation}
N_{\gamma,{\rm init}}=\frac{L_{\rm src}\Delta t}{h_{\rm p}\nu N_{\rm rays, init}},
\end{equation}
where $L_{\rm src}$ is the source luminosity.

\begin{figure}
\begin{center}
\includegraphics[width=8cm]{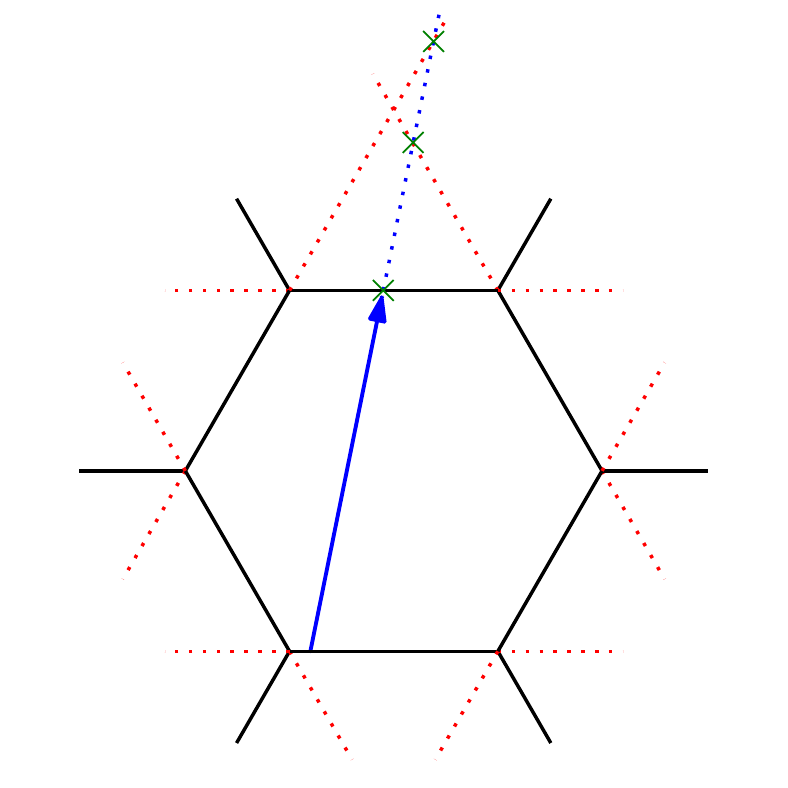}
\caption{Schematic diagram showing the traversal of the rays through the mesh. From the entry point of the ray (solid blue line) at the bottom left of the cell, the intersections (green crosses) with all Voronoi edges in the direction of the ray are computed. The extensions of the edges and the ray are highlighted by the dotted lines. The intersection with the smallest distance to the starting point of the ray is used as the exit point of the current cell and the entry point for the next cell. In three dimensions, the intersections with the Voronoi faces instead of the edges are computed.}
\label{fig_vor}
\end{center}
\end{figure}

The structure holding the ray data is stored on the {\sc MPI} task associated with the ray. The corresponding task is found by using the mesh-generating point that is closest to the initial position of the ray. After the rays have been initialized, they are traversed until they reach the edge of the local computational domain. A global communication step then distributes the rays to their new domains, where the ray walk is continued. Individual rays may be terminated once they reach the edge of the computational domain, or $N_\gamma$ falls below $f_{\rm term}N_{\gamma, {\rm init}}$, where $f_{\rm term}$ may be set to a non-zero value in order to reduce the computational cost of the ray tracing. The traversal of the rays exploits a dynamically updated neighbour list, which for a given cell contains the indices of all neighbouring cells. In two dimensions, the next cell along a ray is found by locating the Voronoi edge that it crosses. This is done by computing the intersections of the Voronoi edges with the ray, and using the intersection that has the smallest distance to the starting point of the ray (see Fig.~\ref{fig_vor}). In three dimensions, the intersections with the Voronoi faces are instead computed. The distance $\Delta l_i$ to the next cell is returned and used for the computation of the optical depth. This step involves only a few arithmetic operations, and is therefore relatively inexpensive \citep{jonsson06}. As the rays are walked, the photon numbers are updated according to:
\begin{equation}
N_{\gamma, i+1}=N_{\gamma, i}\exp{\left(-\tau_i\right)},
\end{equation}
where $i$ denotes the current cell, and $\tau_i=\alpha_i\Delta l_i=\sigma_in_i\Delta l_i$. The heating rate of the cell due to the absorption of $\Delta N_{\gamma,i}=N_{\gamma, i+1}-N_{\gamma, i}$ photons of energy $h_{\rm p}\nu$ is given by
\begin{equation}
\Gamma_{{\rm rad},i}=\frac{\Delta N_{\gamma,i}h_{\rm p}\nu}{V_i\Delta t}
\end{equation}
where $V_i$ is the volume of the cell. The scheme may also be used for multiline, multifrequency radiation transport. In this case, the number of photons in each ray is distributed to $N_{\rm bins}=N_{\rm lines}N_\nu$ bins, and the absorption coefficient becomes a function of line and frequency.

Next to a constant angular resolution, the scheme also allows for spatially adaptive splitting of rays by exploiting the recursive nature of the {\sc healpix} algorithm \citep{aw02}. The parameter $N_{\rm rpc}$ controls the average number of rays per cell by comparing the opening angle associated with a ray, $A_{\rm ray}=4\pi r^2/(12\times 4^l)$, where $r$ is the distance from the source, to an estimate of the cell area, $A_{\rm cell}=\pi h^2$, where $h$ relates the approximate size of a cell to its volume: $h=(3V/4\pi)^{1/3}$. A parent ray is split into four child rays if $A_{\rm cell}<N_{\rm rpc}A_{\rm ray}$, in which case the photons are distributed evenly among the new rays. The scheme also includes a tool for logging and storing the absorption profiles of a selection of rays using a list of unique ray IDs. In addition, the photon escape fraction for each source is computed and stored in the simulation output.

\subsection{H$_2$ line transfer}

For H$_2$ line transfer, the relevant emissivities and cross-sections must be included. Since the line widths are much smaller than the separation of the individual lines in frequency space, they may be treated independently. The emission rate per H$_2$ molecule in each line is given by equation~\ref{eq_eps_h2}. The initial number of photons in cell $i$ due to the transition ${\rm u}\rightarrow {\rm l}$ is therefore given by
\begin{equation}
N_{\gamma,i,{\rm ul,init}}=\frac{\epsilon_{{\rm LTE},i,{\rm u}l}n_{{\rm H_2},i}V_i\Delta t}{h\nu_{\rm ul}N_{\rm rays, init}}.
\end{equation}
In the clouds considered here, the dominant broadening mechanism is thermal Doppler broadening. The number of photons in each line is distributed to $N_\nu$ frequency bins with a maximum displacement of $f_\nu\Delta\nu_{\rm D,ul,init}$ around $\nu_{\rm ul}$, where $\Delta\nu_{\rm D,ul,init}$ is the Doppler width of the emitting cell (given by equation~\ref{eq_nud}), and $f_\nu=3$ yields sufficient accuracy. The positions of the frequency bins, denoted by the index $n$, are given by
\begin{equation}
\nu_{{\rm ul},n}=\nu_{\rm ul}+p_n\Delta\nu_{\rm D,ul,init},
\end{equation}
where
\begin{equation}
p_n=f_\nu \left(\frac{2n+1}{N_\nu}-1\right).
\end{equation}
Finally, the number of photons in each bin is given by
\begin{equation}
N_{\gamma,i,{\rm ul},n,{\rm init}}=q_nN_{\gamma,i,{\rm ul,init}},
\end{equation}
where
\begin{equation}
q_n=K\exp{\left(-p_n^2\right)},
\end{equation}
and $K$ is a normalization constant:
\begin{equation}
\frac{1}{K}=\sum_{n=0}^{N_\nu -1}\exp{\left(-p_n^2\right)}.
\end{equation}

The frequency-dependent cross-section is given by equation~\ref{eq_cross}, with the exception that the frequency in the frame of the cell under consideration is shifted with respect to $\nu_n$:
\begin{equation}
\nu_s=\nu_n\left(1-v_{\rm rel}/c\right),
\end{equation}
where $v_{\rm rel}$ is the relative velocity between cell $i$ and the source cell projected along the ray. Based on the positions of the frequency bins, the cross-section may be written as
\begin{equation}
\sigma_{i,{\rm ul},n}=\sigma_{i,{\rm ul}}\exp{\left(-s_n^2\right)},
\end{equation}
where
\begin{equation}
s_n=\frac{\left(\nu_{\rm ul}+p_n\Delta\nu_{\rm D,ul,init}\right)\left(1-v_{\rm rel}/c\right)-\nu_{\rm ul}}{\Delta\nu_{\rm D,ul}}.
\end{equation}
This equation may be simplified to
\begin{equation}
s_n=p_nr-v_{\rm rel}/v_{\rm th},
\end{equation}
where $r=\Delta\nu_{\rm D,ul,init}/\Delta\nu_{\rm D,ul}$. The term $p_nrv_{\rm rel}/c$ may be omitted, since it is much smaller than the other terms for $v_{\rm rel}\ll c$. The cross-section may differ from the original cross-section due to variations in the thermal Doppler width and the relative velocity of the cells along the rays. The first effect is symmetric around $\nu_{\rm ul}$, while the second may cause a shift in the peak of the cross-section to lower or higher frequencies. Taken together, the optical depth for a given cell, line, and frequency may be written as
\begin{equation}
\tau_{i,{\rm ul},n}=\sigma_{{\rm eff},i,{\rm ul}}\exp{\left(-s_n^2\right)}n_{{\rm H_2}, i}\Delta l_i,
\end{equation}
where $\sigma_{{\rm eff},i,{\rm ul}}$ is given by equation~\ref{eq_cross_eff} and is tabulated as a function of temperature, and $s_n$ is trivially computed from the tabulated $p_n$, the square root of the ratio of the Doppler broadening parameters, and the relative velocity. This minimizes the computational overhead for the $N_{\rm bins}$ computations of the optical depth per cell.

If not all transitions for the H$_2$ energy levels described in Section~\ref{sec_chem} are used, the specific cooling rates and cross-sections are tabulated only for the most luminous $N_{\rm lines}$ lines within each temperature bin. As a result, each cell emanates a different set of lines, and the cross-sections must be adjusted accordingly. This may substantially reduce the computational cost of the ray tracing, while only slightly reducing its accuracy. For example, the simulation described in Section~\ref{sec_sim} uses only $32$ lines, which results in an accuracy of $\simeq 2$ per cent, but reduces the number of opacity calculations and the memory required for the storage of the rays by nearly an order of magnitude.

\subsection{Performance}

On a modern computing core, the ray tracing is able to walk approximately one million cells per second. If more than one opacity bin per cell is used, the performance is approximately reduced by $f_{\rm red}=1/N_{\rm bins}^\beta$, where $\beta\la 1$. Assuming optimal parallelization and no communication cost, the wall-clock time required to complete the ray walk is approximately given by
\begin{equation}
t_{\rm ray}=f_{\rm red}\frac{N_{\rm cells}^{4/3}N_{\rm rays, init}}{N_{\rm cps}N_{\rm cores}},
\end{equation}
where $N_{\rm cps}\simeq 10^6\,{\rm s}^{-1}$ is the number of cells walked per second, and $N_{\rm cores}$ the number of cores. For example, the simulation described in Section~\ref{sec_sim} uses $N_{\rm cells}\simeq 10^7$ and $N_{\rm rays, init}=48$, with $f_{\rm red}\simeq 5$ for this configuration. On 1024 cores, a ray-tracing step should therefore take about $500\,{\rm s}$. However, this theoretical peak performance neglects the significant computational overhead caused by the communication of the rays. The memory required to store the energy of the rays in single precision is $4N_{\rm cells}N_{\rm rays, init}N_{\rm bins}\,{\rm bytes}$, which amounts to $\simeq 0.5\,{\rm TB}$ in the above example. For each communication step, this amount of memory needs to be exchanged between tasks. We have found that this bottleneck typically reduces the performance by a factor of $2$--$3$. Another problem is the significant imbalance in the number of rays stored on the tasks. Since the domain decomposition is optimized for spatial proximity, while in centrally concentrated gas configurations most rays propagate from the centre to the edge of the computational domain, a significant imbalance accumulates as the rays are walked. On average, the ray numbers differ by a factor of $2$--$3$, which leads to a similar reduction in performance. Overall, the performance is therefore reduced by a factor of a few compared to the theoretical maximum. Since the communication of the rays takes up a large portion of the total computational cost, it is not yet worthwhile to optimize the scheme for use on graphics processing units or coprocessors.

The relatively high performance of the ray walk is achieved by various optimizations. First, {\sc arepo} orders the cells in memory based on two nested Peano--Hilbert curves, each consisting of $2^{21}$ elements per dimension. As the rays are walked, cells that are close to each other in space are therefore also close in memory. This facilitates a quick look-up of the properties of neighbouring cells. Second, most quantities required for the computation of the optical depth are tabulated. Since neighbouring cells have similar opacities, the variables required to compute the opacities are thus likely already stored in the cache. Finally, for simulations run on modern Intel cores, the Intel compiler generates code that fully exploits the Advanced Vector Extensions (AVX) instruction set. In this case up to eight single-precision operations may be performed simultaneously. With these optimizations, the calculation of the optical depth in the simulation discussed in Section~\ref{sec_sim} with $256$ opacity bins per cell is only a factor of $\simeq 5$ slower than the case where one bin is used.

In order to reduce the cost of the communication in runs with large core counts, the ray tracing may be used with a hybrid shared/distributed memory scheme. In this case, the rays are distributed to multiple {\sc OpenMP} threads, and each thread processes its own list of rays. Since each thread is assigned nearly the same amount of rays, and the average computational cost associated with a ray does not fluctuate by much, the scheme achieves near ideal work balance. The use of multithreading is particularly advantageous for simulations with comparatively small cell counts per core. In the above example, the computational cost may be significantly reduced if the simulation is run with four tasks and four threads instead of $16$ tasks and one thread per node.

\begin{figure*}
\begin{center}
\includegraphics[width=17cm]{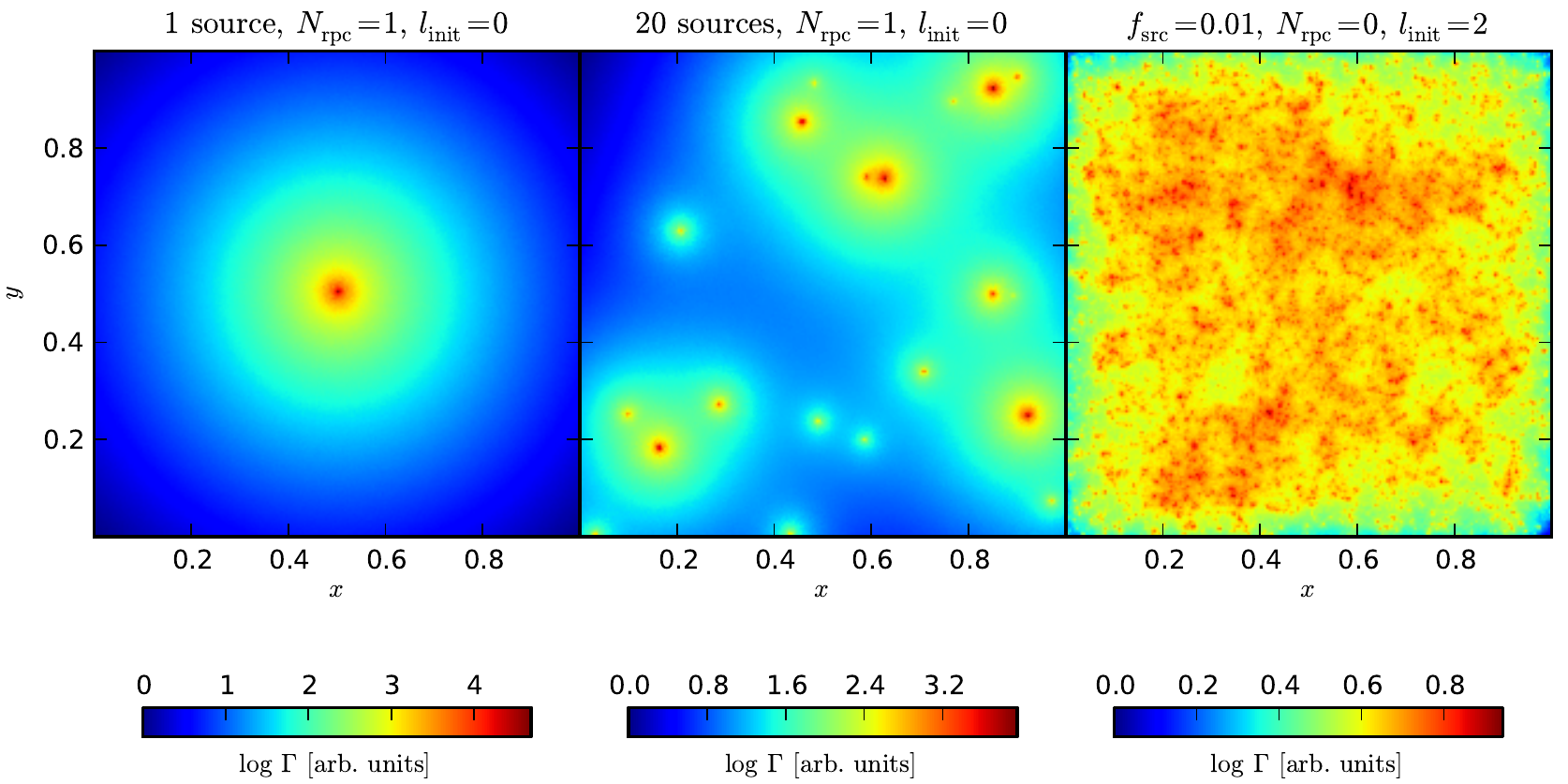}
\caption{Test calculations that apply the ray-tracing scheme to various source configurations in a uniform density box. The heating rate averaged along the line of sight is colour coded in arbitrary units. The first two panels show the radiation from isolated point sources, where the rays are adaptively refined such that each source maintains approximately one ray per cell. The panel on the right-hand side shows diffuse radiation from a significant fraction of all cells in the computational domain with a fixed angular resolution. The heating rate is isotropic around the individual sources, and shows the expected rapid decay with increasing distance.}
\label{fig_tray_img}
\end{center}
\end{figure*}

\subsection{Test calculations} \label{sec_test}

We here use a series of idealized test calculations to investigate the accuracy and convergence of the ray-tracing scheme. The ray tracing may be run in a distinct mode in which $N_{\rm src}$ cells are randomly chosen as sources, with luminosities that vary randomly within a pre-defined range. The attenuation along the rays is modelled with a spatially homogeneous absorption coefficient $\alpha$. In the limit of infinitesimally small volumes, the radiative heating rate of cell $i$ is given by
\begin{equation}
\Gamma_{{\rm rad},i}=\frac{\alpha}{4\pi}\sum_j L_{{\rm src},j}\frac{\exp{\left(-\alpha r_{ij}\right)}}{r_{ij}^2},
\end{equation}
where the sum extends over all sources indexed by $j$, $L_{{\rm src},j}$ denotes the luminosities of the sources, and $r_{ij}$ their separation from cell $i$. We note that the heating rate derived from the absorption of the radiation in the simulations is subject to a small systematic error due to the finite lengths of the cells in the direction of the rays. By varying the resolution of the mesh, we have verified that in the test cases presented below this error is small compared to the error that arises from the limited angular resolution of the ray tracing.

In the test calculations, we use uniform density boxes with $100^3$ cells. The mesh-generating points are displaced from a cubical lattice by $\Delta x,\Delta y,\Delta z=0.01f_{\rm rnd}\Delta a$, where $f_{\rm rnd}$ is a random number in the range $\left\{0,1\right\}$ and $\Delta a$ is the grid spacing. The luminosities of the sources are varied by two orders of magnitude, and the absorption coefficient is given by $\alpha=5$ in inverse units of $r$. We perform a single ray-tracing step without any dynamical evolution, such that all gas properties except the heating rate remain unaffected. In Fig.~\ref{fig_tray_img}, we show the heating rate averaged along one of the axis of the simulation box for one, $20$, and $10^4$ sources. The first two cases demonstrate the adaptative resolution of the ray tracing using $l_{\rm init}=0$ and $N_{\rm rpc}>0$, which is typically used for isolated point sources. The third simulation demonstrates an application to diffuse radiation, where $f_{\rm src}\le 1$, $l_{\rm init}\ge 0$, and $N_{\rm rpc}=0$. Evidently, the heating rate is isotropic around the individual sources, and shows the expected rapid decay with increasing distance.

\begin{figure}
\begin{center}
\includegraphics[width=8cm]{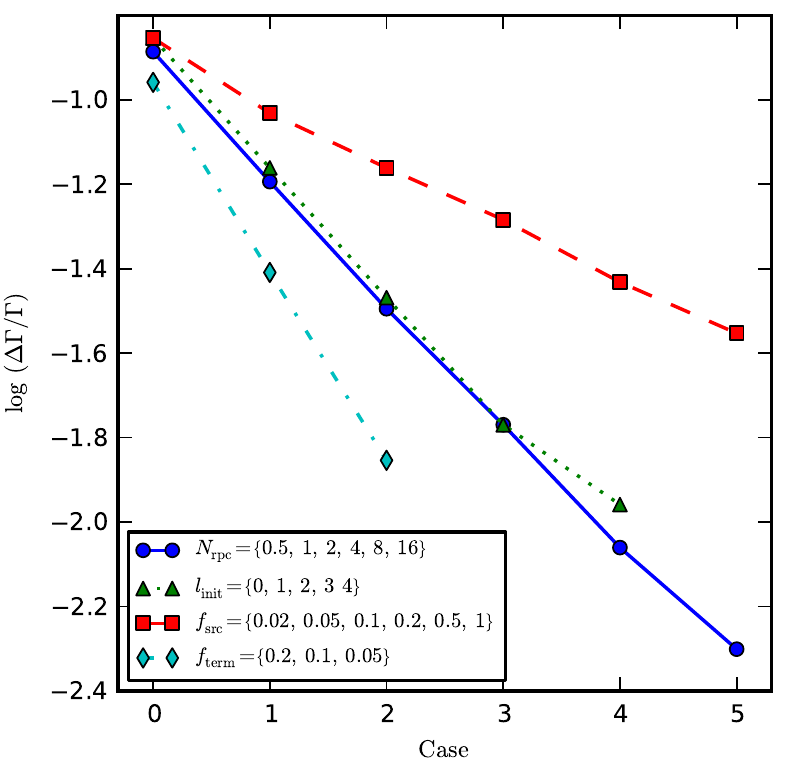}
\caption{The convergence rate for the test calculations shown in Fig.~\ref{fig_tray_img}. The relative error is computed by comparing the analytic heating rate with that obtained with the ray tracing, averaged over all cells in the computational domain. The solid blue line shows the error as a function of the number of rays per cell for the test simulation shown in the middle panel of Fig.~\ref{fig_tray_img}. The other profiles show the convergence rate for diffuse radiation, corresponding to the panel on the right-hand side of Fig.\ref{fig_tray_img}. The green dotted line denotes the case where the angular resolution is varied, while the other parameters are kept fixed. In this case, the error decreases nearly linearly with increasing {\sc healpix} level. The dashed red line shows the error as a function of the source fraction $f_{\rm src}$, and the dot--dashed cyan line as a function of the ray termination fraction $f_{\rm term}$. Reducing the source fraction usually results in a large error, while an increased termination fraction may only slightly degrade the accuracy, but substantially reduce the computational cost of the ray tracing.}
\label{fig_tray_err}
\end{center}
\end{figure}

A more quantitative analysis of the test calculations is shown in Fig.~\ref{fig_tray_err}. The relative error is computed by comparing the analytic heating rate to that obtained with the ray tracing, averaged over all cells in the computational domain. The solid blue line shows the error as a function of $N_{\rm rpc}$ for $20$ sources, corresponding to the test case shown in the middle panel of Fig.~\ref{fig_tray_img}. As the number of rays per cell is increased, the heating rate converges nearly linearly, with an error of about $7$ per cent for $N_{\rm rpc}=1$. The other three profiles show the convergence for diffuse radiation, similar to the case shown in the right-hand panel of Fig.~\ref{fig_tray_img}. In the first case, corresponding to the green dotted line, the angular resolution is varied while the other parameters are kept fixed, with $f_{\rm src}=0.1$ and $f_{\rm term}=0$. Similar to the case where $N_{\rm rpc}$ is varied, the heating rate shows approximately linear convergence. However, this is somewhat misleading, since the change in angular resolution between two {\sc healpix} levels is a factor of $4$, while $N_{\rm rpc}$ only varies by a factor of $2$. The convergence rate is therefore sublinear. The dashed red line shows the error as a function of the source fraction for $l_{\rm init}=1$ and $f_{\rm term}=0$. The error introduced by the fixed angular resolution becomes smaller as the number of sources is increased, but at a slower rate. Finally, the dot--dashed cyan line shows the error as a function of $f_{\rm term}$ for $l_{\rm init}=4$ and $f_{\rm src}=0.1$. In this case, the solution rapidly converges to the $l_{\rm init}=4$ case. For large optical depths, using $f_{\rm term}>0$ may significantly reduce the computational cost of the simulation, while only slightly degrading the accuracy of the ray tracing. On the other hand, reducing $f_{\rm src}$ typically results in a large error.

These calculations show that the ray-tracing scheme is capable of reproducing known analytic solutions. The convergence rate depends on the physical problem under consideration. It is therefore essential to investigate the dependence of the error on the parameters of the ray tracing, and then choose the parameters such that the desired accuracy is achieved. For isolated point sources, the relevant parameters are $N_{\rm rpc}$ and $f_{\rm term}$, while for diffuse radiation they are $l_{\rm init}$, $f_{\rm src}$, and $f_{\rm term}$. If multiple lines and frequency bins are used, the additional parameters $N_{\rm lines}$ and $N_\nu$ must be specified.

\section{Simulations} \label{sec_sim}

We here describe the set-up of the main simulations, which are performed with the moving-mesh code {\sc arepo} \citep{springel10a}. Because of the very high computational cost of the H$_2$ line transfer, we perform only one simulation where the ray tracing is fully coupled to the hydrodynamics. For the resolution study and other comparisons, we use the ray tracing as a post-processing tool on the output of the main simulation.

\subsection{Dark matter simulations}

The cosmological parameters used to initialize the simulations are obtained from the CMB measurements by the {\it Wilkinson Microwave Anisotropy Probe} \citep{komatsu09}. These are the matter density $\Omega_{\rm m}=0.27$, baryon density $\Omega_{\rm b}=0.046$, Hubble parameter $h_{\rm H}=0.7$, spectral index $n_{\rm s}=0.96$, and normalization $\sigma_8=0.81$. The matter power spectrum is evolved forward in time until $z=99$, after which the Zel'dovich approximation is used to determine the initial displacements of the DM particles on a cubical lattice. We use a box with a side length of $1\,{\rm Mpc}$ (comoving), $512^3$ particles of mass $\simeq 272\,{\rm M}_\odot$, and a gravitational softening length of $\simeq 98\,{\rm pc}$ (comoving), which corresponds to $5$ per cent of the initial mean interparticle separation. The DM simulation is evolved until the first halo grows to a virial mass of $M_{\rm vir}=5\times 10^5\,{\rm M}_\odot$, which is evaluated by an on-the-fly friends-of-friends algorithm \citep{springel01}.

\subsection{Extraction and resimulations}

Once the target halo has been found, the simulation is centred on the halo and reinitialized with higher resolution. The DM particles in the target halo as well as a sufficiently large boundary region around it are traced back to their initial positions, which yield the Lagrangian volume out of which the halo formed. In this region, each low-resolution particle is replaced by $16^3$ high-resolution particles, and augmented with additional small-scale power. Gas cells are placed next to the DM particles, and the relative displacements are set to half of the initial mean interparticle separation, using a mass ratio $M_{\rm gas}/M_{\rm dm}=\Omega_{\rm b}/(\Omega_{\rm m}-\Omega_{\rm b})$. Outside of the target region, the resolution is decreased by factors of $8$ down to an effective resolution of $32^3$, subject to the constraint that the accuracy of the gravitational tidal field around the halo is preserved. The DM particle and gas cell masses in the high-resolution region are $\simeq 0.05$ and $\simeq 0.01\,{\rm M}_\odot$, respectively, and the gravitational softening length is $\simeq 6\,{\rm pc}$ (comoving). The cosmological resimulations are evolved to a density of $n_{\rm H}=10^9\,{\rm cm}^{-3}$, where the maximum dynamic range that can be simulated efficiently is reached \citep{gsb13}. The central $1\,{\rm pc}$ of the box is cut out and reinitialized using inflow/outflow boundary conditions, and evolved to a density of $n_{\rm H}=10^{15}\,{\rm cm}^{-3}$.

\subsection{Refinement}

Following \citet{gsb13}, cells are refined if $h>\lambda_{{\rm J},200}/N_{\rm J}$, where $N_{\rm J}$ is the desired number of cells per Jeans length, and $\lambda_{{\rm J},200}$ is the Jeans length evaluated at $200\,{\rm K}$:
\begin{equation}
\lambda_{{\rm J},200}=0.6\,{\rm pc}\left(\frac{n_{\rm H}}{10^4\,{\rm cm}^{-3}}\right)^{-1/2}.
\end{equation}
To ensure an adequate resolution of the turbulent cascade, we use $32$ cells per Jeans length. For this choice, the number of cells increases to $\simeq 2\times 10^7$ at $n_{\rm H}=10^{15}\,{\rm cm}^{-3}$. Using an even higher resolution would increase the computational cost of the ray tracing to a level where it is no longer possible to maintain the desired accuracy. Next to the Jeans refinement, we also refine cells if their mass increases to more than twice their initial mass.

\subsection{H$_2$ line emission}
In the resimulations, we use three different prescriptions for the optically thick H$_2$ line cooling rate. In the first case, denoted by MH-Fit, the density-dependent fitting function discussed in Section~\ref{sec_fit} is used. In the second case, denoted by MH-Sob, we use the Sobolev method described in Section~\ref{sec_sob}. Finally, in MH-Ray we use the ray-tracing scheme described in Sections~\ref{sec_ray} and \ref{sec_rad}. Since a relatively small Courant factor of $0.2$ must be used for numerical stability of the hydrodynamic solver, the ray tracing is performed only every fifth time step. The simulations are run on nodes with two Intel Sandy bridge processors that have eight cores each, and $32\,{\rm GB}$ of memory. For MH-Fit and MH-Sob, $16$ nodes with $16$ {\sc MPI} tasks per node are used, while for MH-Ray $64$ nodes with four tasks and four threads per node are used. The computational cost of the simulation is dominated by the ray tracing, which takes about $\simeq 2000\,{\rm s}$ for each step. Since approximately $10^4$ time steps are necessary to evolve the simulations to a density of $n_{\rm H}\simeq 10^{15}\,{\rm cm}^{-3}$, the total wall-clock time is $1$--$2$ months, which is equivalent to about one million CPU hours.

\section{Results} \label{sec_res}

We here present the resolution study that is used to gauge the parameters of the ray tracing, followed by a discussion of the results of the simulations.

\subsection{Resolution study}

\begin{figure}
\begin{center}
\includegraphics[width=8cm]{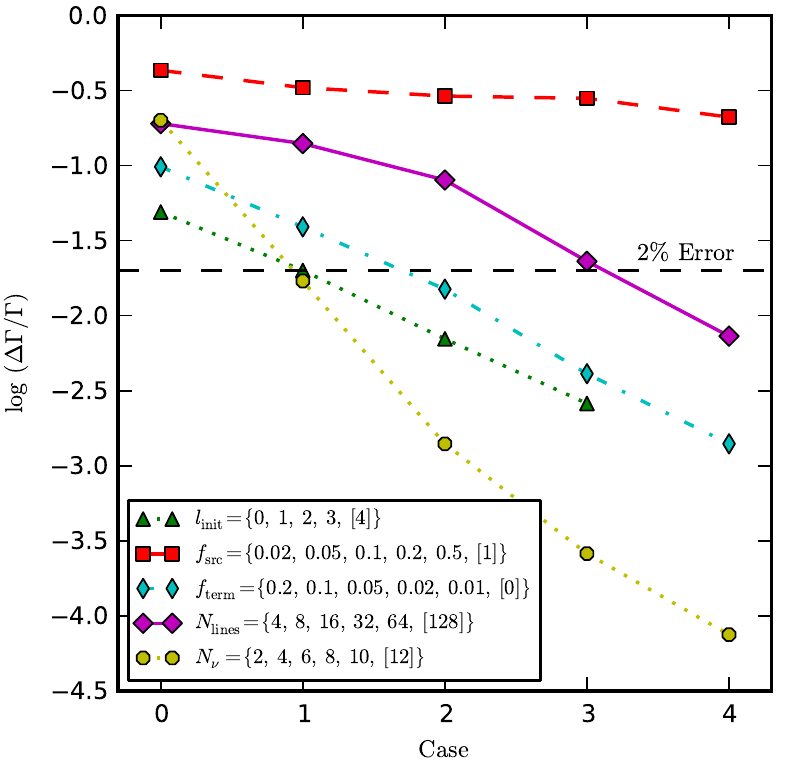}
\caption{The error in the heating rate as a function of the various parameters used in the ray tracing for the H$_2$ line emission. The relevant parameters are the initial {\sc healpix} level $l_{\rm init}$, the source fraction $f_{\rm src}$, the ray termination fraction $f_{\rm term}$, the number of lines $N_{\rm lines}$, and the number of frequency bins $N_\nu$. The error is determined by comparing the heating rate for a given parameter choice to the case where the parameter under consideration is set to the most accurate value, denoted by the number in the square brackets. With the exception of the source fraction, the heating rate converges relatively well. Assuming the errors associated with the variation of the individual parameters are uncorrelated, the desired accuracy can be achieved by choosing the parameters appropriately. In the fully coupled radiation hydrodynamics simulations, we use $l_{\rm init}=1$, $f_{\rm src}=1$, $f_{\rm term}=0.05$, $N_{\rm lines}=32$, and $N_{\nu}=8$. This results in an overall accuracy of $\simeq 5$ per cent.}
\label{fig_mh_err}
\end{center}
\end{figure}

The parameters required to obtain the desired accuracy in the heating rate may be gauged with the resolution study shown in Fig.~\ref{fig_mh_err}. The error is determined by comparing the heating rate for a certain parameter choice to the case where the parameter under consideration is set to the most accurate value. All other parameters are set to their default values. The calculations are performed using the final snapshot of MH-Ray, and include gas at all densities relevant for optically thick H$_2$ line emission. As opposed to the test calculations in Section~\ref{sec_test}, the correct solution is not known. The true error may therefore be somewhat larger than that shown in Fig.~\ref{fig_mh_err}. However, in almost all cases the error decreases at a nearly constant rate, which indicates convergence. Assuming the errors associated with the variation of the individual parameters are uncorrelated, the total error is equal to the sum of the individual errors.

The angular resolution is varied by increasing $l_{\rm init}$ from $0$ to $4$. We choose $l_{\rm init}=1$ as the default value, since this yields an error of approximately $2$ per cent. The source fraction is varied between $0.02$ and 1. However, since the resulting error is very large and shows poor convergence, we choose $f_{\rm src}=1$ and thereby eliminate this source of uncertainty. This large error is partly due to the high optical depth of the gas at the highest densities, where a ray may be terminated in the same cell it was initialized in. The parameter $f_{\rm term}$ is varied between $0.2$ and $0$, and the resulting heating rate converges relatively quickly. We use $f_{\rm term}=0.05$, which yields slightly better than $2$ per cent accuracy and leads to a significant increase in performance. The number of H$_2$ lines is varied between $4$ and $128$, and displays a substantial increase in accuracy for $N_{\rm lines}>16$. We therefore use $N_{\rm lines}=32$, which yields an accuracy of approximately $2$ per cent. Finally, the number of frequency bins is varied between $2$ and $12$. The error decreases very rapidly with increasing resolution, as the error falls below $0.01$ per cent already for $N_\nu=12$. Even though $4$--$6$ frequency bins would be sufficient, we use $N_\nu=8$ and thereby effectively eliminate this source of error. For $l_{\rm init}=1$, $f_{\rm src}=1$, $f_{\rm term}=0.05$, $N_{\rm lines}=32$, and $N_{\nu}=8$, the overall error is therefore approximately $5$ per cent.

\subsection{Cloud properties}

\begin{figure*}
\begin{center}
\includegraphics[width=17cm]{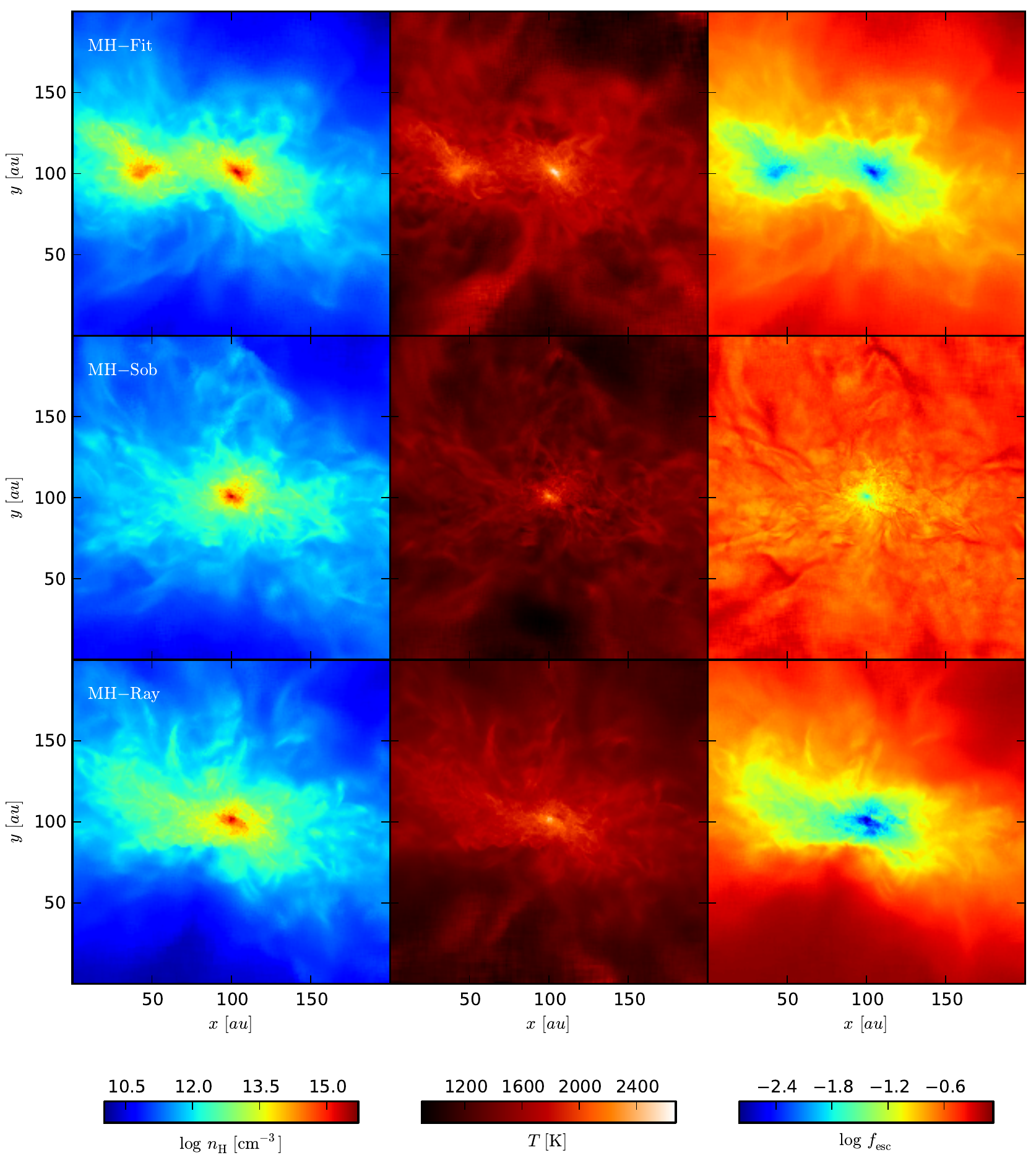}
\caption{The gas clouds that form in the central $200\,{\rm au}$ of the minihalo, shown when the central density exceeds $n_{\rm H}=10^{15}\,{\rm cm}^{-3}$. From left to right: number density of hydrogen nuclei, temperature, and escape fraction averaged along the line of sight. From top to bottom: simulations using differing prescriptions for the optically thick H$_2$ line cooling rate: a density-dependent fitting function \citep[MH-Fit;][]{ra04}, the Sobolev method \citep[MH-Sob;][]{yoshida06b}, and a self-consistent radiation hydrodynamics simulation (MH-Ray; this work). In MH-Sob and MH-Ray, a centrally concentrated cloud has formed, while in MH-Fit the cloud has fragmented into two distinct clumps. This is similar to the subfragmentation found in \citet{tao09} and \citet{gsb13}. The cloud that forms in MH-Sob is slightly more concentrated and cooler than in the other two cases. The escape fraction at the centre of the box drops to $\simeq 2\times 10^{-3}$ in MH-Ray and MH-Fit, while in MH-Sob the escape fraction has a very different spatial pattern and only drops to $\simeq 0.03$.}
\label{fig_mh_img}
\end{center}
\end{figure*}

The properties of the central gas cloud in the three simulations are shown in Fig.~\ref{fig_mh_img}. The individual columns show the number density of hydrogen nuclei, temperature, and escape fraction in the central $200\,{\rm au}$ of the box at the final output time. In agreement with previous mesh-based studies, the gas clouds are centrally concentrated and have a filamentary morphology, which is indicative of turbulence \citep{tao09, gsb13}. In MH-Sob and MH-Ray, a single clump has formed, while in MH-Fit the cloud has fragmented into two distinct clumps. Similar subfragmentation was also found in the simulations of \citet{tao09} and \citet{gsb13}. The cloud that forms in MH-Sob is slightly more concentrated than in the other two cases. The temperature increases much more gradually than the density, and ranges from about $1000\,{\rm K}$ at the edge of the cloud to $\simeq 2500\,{\rm K}$ at the centre. In MH-Sob, the cloud is somewhat cooler than in MH-Fit and MH-Ray. The escape fraction decreases to nearly $2\times 10^{-3}$ at the centre of the cloud in MH-Ray and MH-Fit, while in MH-Sob it decreases to only $\simeq 0.03$. The spatial pattern of the escape fraction in MH-Sob is also very different from that in MH-Fit and MH-Ray.

\begin{figure*}
\begin{center}
\includegraphics[width=16cm]{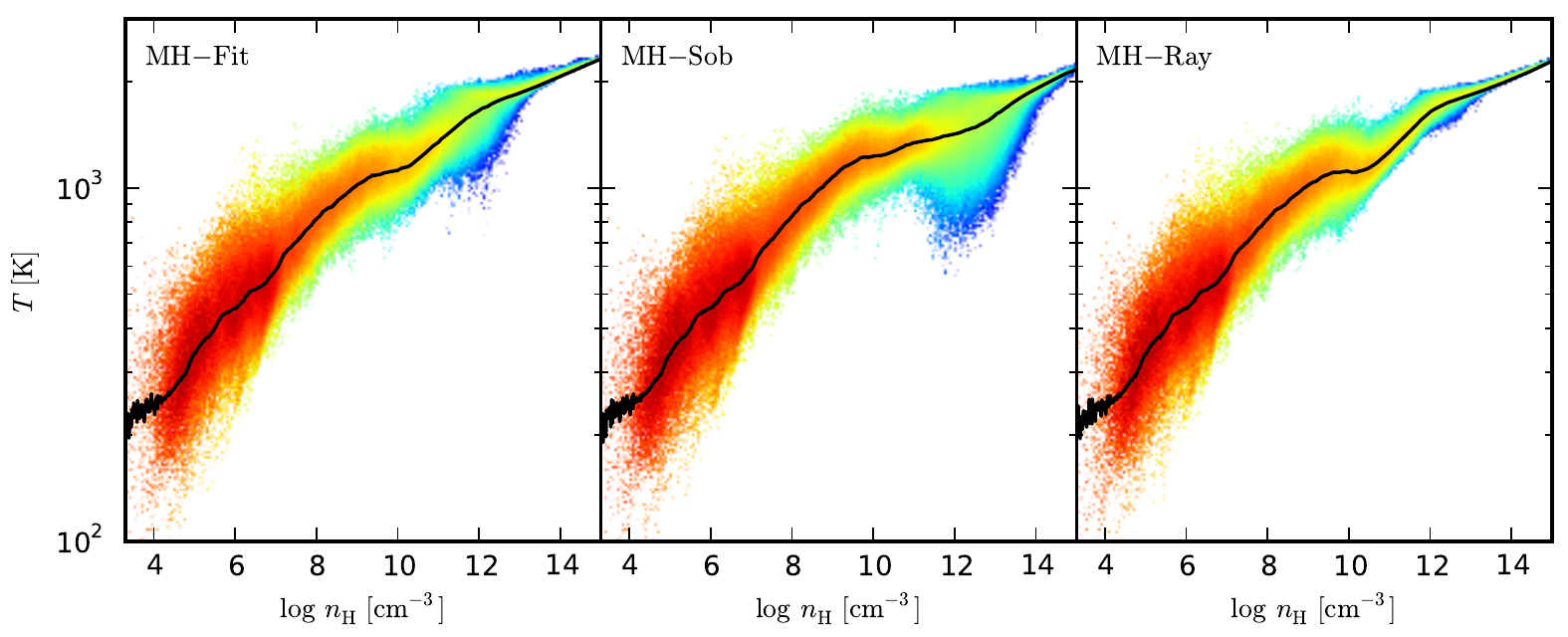}
\caption{Density and temperature distribution of the gas. The logarithm of the mass per
bin over the total mass in the computational domain is colour coded from blue (lowest) to red (highest). The solid black line shows the mass-weighted average values. In MH-Sob, the average temperature at a density of $10^{12}\,{\rm cm}^{-3}$ is somewhat lower than in the other simulations, but at $n_{\rm H}=10^{15}\,{\rm cm}^{-3}$ all three simulations converge to $T\ga 2100\,{\rm K}$. In MH-Sob, the temperature at $n_{\rm H}\simeq 10^{12}\,{\rm cm}^{-3}$ varies by a factor of $5$, while in MH-Fit and MH-Ray the dispersion is significantly smaller. The dispersion continues to decrease with increasing density, indicating that the clouds become increasingly spherically symmetric. The pronounced thermal instability found in \citet{gsb13} is thus an artefact of using the Sobolev method to compute the escape fraction.}
\label{fig_mh_pspace}
\end{center}
\end{figure*}

In Fig.~\ref{fig_mh_pspace}, we show the mass-weighted distribution of the gas in density and temperature, as well as the mass-weighted average temperature versus density. In MH-Fit and MH-Ray, the temperature increases to $\simeq 1000\,{\rm K}$ at a density of $10^9\,{\rm cm}^{-3}$, followed by a brief phase of nearly isothermal contraction to $n_{\rm H}\simeq 10^{10}\,{\rm cm}^{-3}$. The temperature then increases more sharply to $\simeq 1600\,{\rm K}$ at $n_{\rm H}=10^{12}\,{\rm cm}^{-3}$, followed by a more gradual increase to $\simeq 2200\,{\rm K}$ at $n_{\rm H}=10^{15}\,{\rm cm}^{-3}$. The thermal evolution of the gas in MH-Sob is somewhat different. At a density of $10^9\,{\rm cm}^{-3}$, the temperature has already increased to $\simeq 1100\,{\rm K}$ instead of $\simeq 1000\,{\rm K}$. The temperature then gradually rises to $\simeq 1600\,{\rm K}$ at $n_{\rm H}=10^{13}\,{\rm cm}^{-3}$, followed by a relatively sharp increase to $\simeq 2200\,{\rm K}$ at $n_{\rm H}=10^{15}\,{\rm cm}^{-3}$. The overall distribution of the gas shows more pronounced differences. In MH-Sob, the temperature dispersion at a density of $10^{12}\,{\rm cm}^{-3}$ is significantly larger than in the other simulations, and spans approximately a factor of $5$. This is similar to the results of \citet{gsb13}, where the Sobolev method was used. In MH-Fit, the temperature varies only by a factor of $2$, and in MH-Ray the variation is nearly absent. The dispersion continues to decrease with increasing density, indicating that the cloud becomes increasingly spherically symmetric. The chemothermal instability that operates at these densities is significantly more pronounced in MH-Sob, even though the cloud only fragments in MH-Fit. This indicates that fragmentation during the initial collapse is highly stochastic in nature.

The radial profiles of the number density of hydrogen nuclei, temperature, H$_2$ fraction, escape fraction, radial velocity over sound speed, and root-mean-squared density contrast versus radius are shown in Fig.~\ref{fig_mh_rad}. The latter is given by
\begin{equation}
\sigma_\delta=\sqrt{\sum_i \frac{m_i}{M_{\rm bin}}\left(\frac{\rho_i-{\rho_{\rm bin}}}{\rho_{\rm bin}}\right)^2},
\end{equation}
where the sum extends over all cells contributing to a radial bin, $i$ denotes the cell index, $m_i$ the mass, $\rho_i$ the density, $M_{\rm bin}$ the total mass in the bin, and $\rho_{\rm bin}$ the mass-weighted average density. The density profiles show that the flat core of the central, Jeans-unstable cloud extends to a few au. Outside of the core, the density falls off as approximately $n_{\rm H}\propto r^{-2.2}$, which is expected for an effective adiabatic index of $\gamma_{\rm eff}\simeq 1.1$ \citep{larson69, penston69, on98}. The temperature profiles show the same trends as in Fig.~\ref{fig_mh_pspace}. In MH-Sob, the central temperature is somewhat lower than in MH-Fit and MH-Ray. As a result, the H$_2$ fraction in MH-Sob is $\simeq 0.4$ at $n_{\rm H}=10^{15}\,{\rm cm}^{-3}$, while in MH-Fit and MH-Ray slightly more H$_2$ has been dissociated, with $y_{\rm H_2}\simeq 0.3$. The somewhat lower temperature in MH-Sob also slightly increases the Mach number of the inflow on a scale of $\simeq 100\,{\rm au}$. These differences can be attributed to the much higher escape fraction in MH-Sob than in MH-Fit and MH-Ray, which also indirectly affects the growth of density fluctuations. This is evident from the density contrast shown in the bottom right-hand panel of Fig.~\ref{fig_mh_rad}. The higher escape fraction in MH-Sob results in a softer effective equation of state, which allows individual parcels of gas to become more dense than in MH-Fit and MH-Ray.

\subsection{Escape fraction} \label{sec_fesc}

\begin{figure}
\begin{center}
\includegraphics[width=8cm]{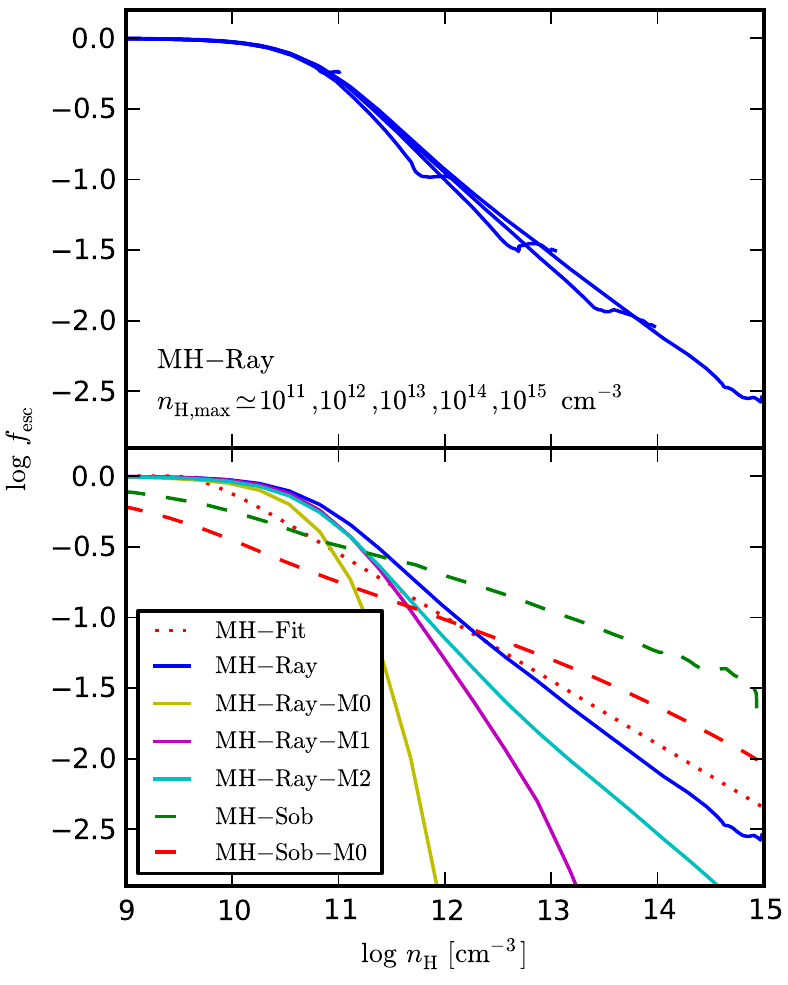}
\caption{Top panel: escape fraction versus number density of hydrogen nuclei for MH-Ray at different peak densities. Bottom panel: escape fraction for MH-Fit (dotted line), two different Sobolev methods (dashed lines), and four different ray-tracing calculations (solid lines). The latter show the influence of various radiative transfer effects. The yellow line shows the escape fraction for a line-averaged grey opacity, the magenta line for the multiline grey opacity, the cyan line for a line and frequency-dependent opacity, and the blue line for the full radiative transfer calculations including variations in the thermal Doppler width and relative velocities. The dashed red line shows the escape fraction using the Sobolev method with the Jeans length instead of the Sobolev length, and the dashed green line denotes the standard Sobolev method. These profiles are discussed in detail in Section~\ref{sec_fesc}. During the initial collapse phase, the escape fraction in MH-Ray agrees relatively well with the fitting function of \citet{ra04}, while for high optical depths the Sobolev method overestimates the escape fraction by more than an order of magnitude.}
\label{fig_mh_fesc}
\end{center}
\end{figure}

The escape fraction at various peak densities is shown in the top panel of Fig.~\ref{fig_mh_fesc}. Apart from the varying sizes of the Jeans-unstable cores, they do not differ much from the final profile. The various processes that operate in MH-Ray can be understood from the bottom panel of Fig.~\ref{fig_mh_fesc}, which shows the escape fraction for three different ray-tracing calculations in addition to MH-Ray, denoted by MH-Ray-M0, MH-Ray-M1, and MH-Ray-M2. In MH-Ray-M0, the line-averaged grey opacity of equation~\ref{eq_cross_avg} is used. In this case, the optical depth is very large, since the lines that emit the most energy also have the highest cross-sections, while lines with a lower cross-section do not contribute substantially to the emission. In MH-Ray-M1, the lines are treated separately using the cross-sections at the centres of the lines. In this case, the energy emitted in the more energetic lines can escape more easily, since the cross-sections are lower, resulting in an escape fraction that is up to two orders of magnitude higher than in the line-averaged case. In MH-Ray-M2, frequency-dependent emission and absorption are taken into account, but deviations in the thermal Doppler width and relative velocities along the rays are neglected. Since the cross-section is much lower in the wings of the lines, the associated energy can escape much more easily, increasing the escape fraction by up to an order of magnitude compared to the grey case. Finally, in MH-Ray all effects are taken into account. In this case, the escape fraction increases by about a factor of $2$ compared to MH-Ray-M2. We have verified that variations in the Doppler width have almost no effect on the escape fraction. The observed difference is therefore due to the Doppler shift induced by relative velocity fluctuations along the rays.

\begin{figure}
\begin{center}
\includegraphics[width=8cm]{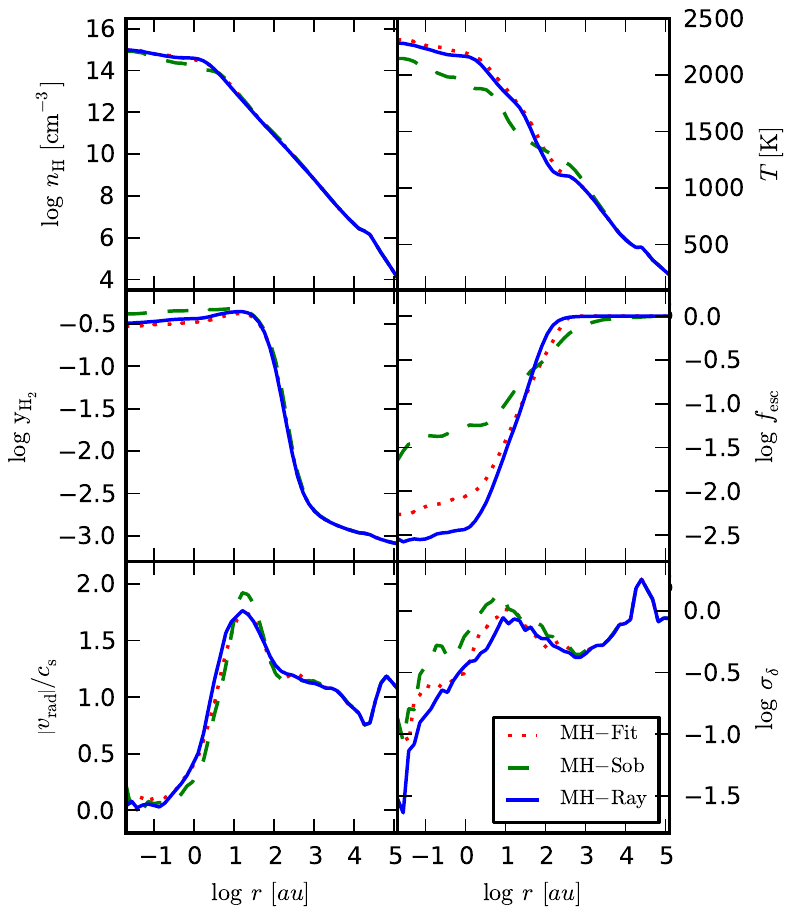}
\caption{From top left to bottom right: radial profiles of the number density of hydrogen nuclei, temperature, H$_2$ fraction, escape fraction, radial velocity over sound speed, and root-mean-squared density contrast. The much higher escape fraction of H$_2$ line emission in MH-Sob results in a somewhat reduced central temperature, an increased H$_2$ fraction, and a higher Mach number for the radial inflow on a scale of $100\,{\rm au}$ compared to MH-Fit and MH-Ray. In addition, due to the softer effective equation of state, the density contrast in MH-Sob is somewhat elevated.}
\label{fig_mh_rad}
\end{center}
\end{figure}

\begin{figure*}
\begin{center}
\includegraphics[width=16cm]{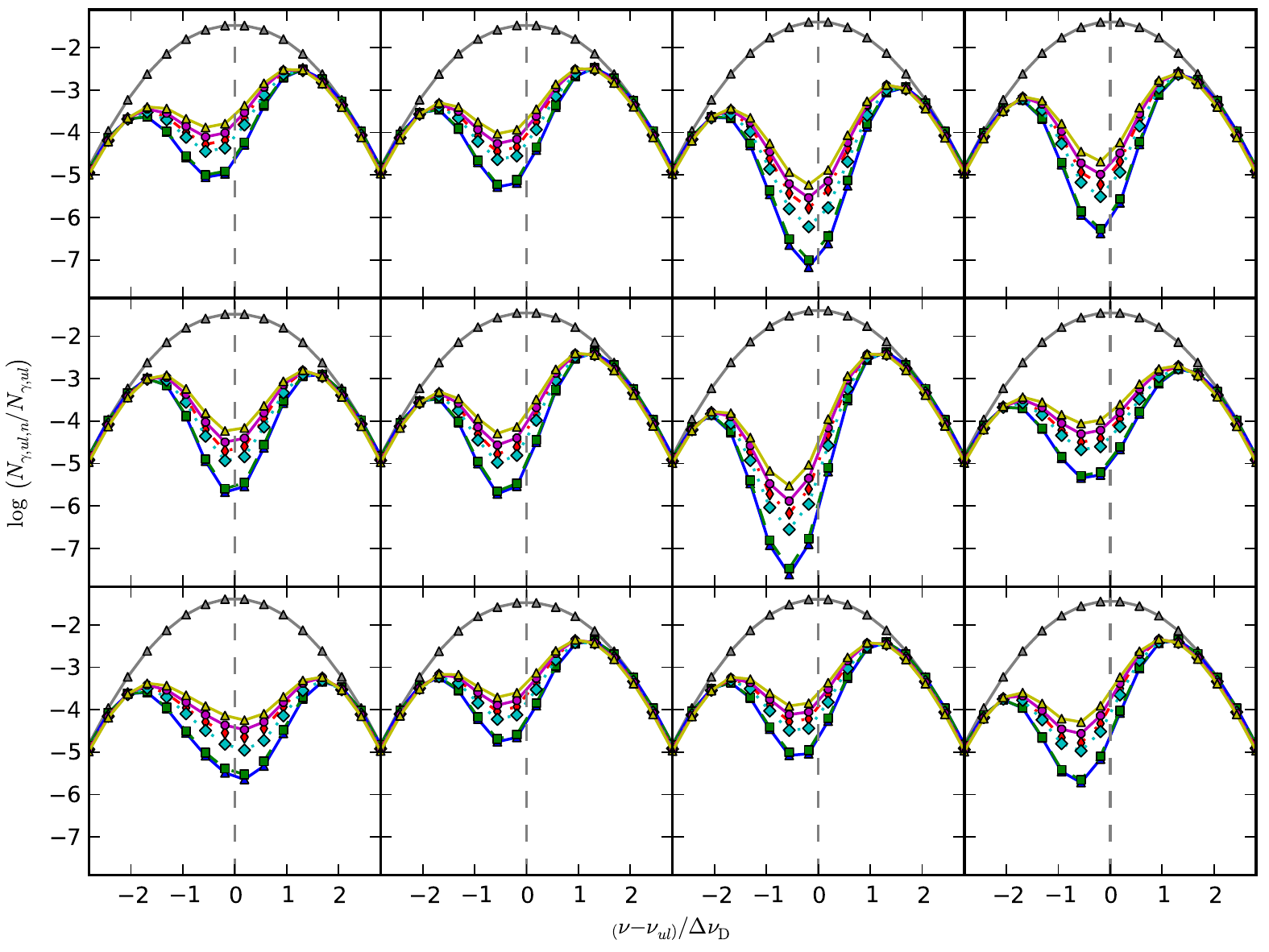}
\caption{Initial and final line profiles for the six most energetic lines in $12$ different rays corresponding to the base {\sc healpix} level, traced from the centre of the cloud to the edge of the computational domain (each panel corresponds to a different ray). The $x$-axis denotes the frequency in units of the thermal Doppler width, and the $y$-axis  denotes the energy in each frequency bin, normalized to the total energy per line. The initial line profiles are represented by the solid grey line, since they are nearly identical. The optical depth is high enough that the centres of the lines are more strongly damped than the wings. In almost all cases, the low-frequency end is more damped, showing that a net gradient in the inflow velocity exists. However, the average shift is only $\simeq\Delta\nu_{\rm D}/2$, implying that relative velocities have only a mild effect on the escape fraction.}
\label{fig_mh_freq}
\end{center}
\end{figure*}

In Fig.~\ref{fig_mh_freq}, we show the initial and final line profiles of the $12$ rays of the base {\sc healpix} level around the densest cell in a ray-tracing calculation with $l_{\rm init}=0$, $N_{\rm lines}=6$, and $N_\nu=16$, using a snapshot when the density first exceeds $n_{\rm H}=10^{12}\,{\rm cm}^{-3}$. Since the initial profiles are so similar, they are represented by the solid grey line. The final profiles are significantly distorted from their initial Gaussian shape due to the strong attenuation along the rays. The various lines show differing amounts of absorption, but in general the centres of the lines are damped more strongly than the wings of the lines. The lines are also attenuated asymmetrically, which is caused by relative velocity fluctuations along the rays. In all but one case, the low-frequency end displays stronger attenuation, which indicates that a net gradient in the inflow velocity exists. However, the average shift is only $\Delta\nu_{\rm D}/2$, showing that relative velocities only mildly affect the escape fraction.

The dashed green line in the top panel of Fig.~\ref{fig_mh_fesc} shows the escape fraction obtained with the Sobolev method. For large optical depths, the escape fraction deviates by more than an order of magnitude from the correct solution. The large error may be attributed to the inherent assumption in equation~\ref{eq_tau} that the velocity gradient that is present in the central, Jeans-unstable cloud extends indefinitely. However, Fig.~\ref{fig_mh_rad} shows that this is not the case. To first order, the infall velocity increases roughly linearly to the sound speed at the Jeans length, and remains constant thereafter. The average velocity fluctuation is thus less than the sound speed. The escape fraction obtained for this idealized case is shown by the dashed red line in Fig.~\ref{fig_mh_fesc}. In this case, denoted by MH-Sob-M0, the Sobolev length is replaced by the Jeans length, which yields a velocity gradient of $c_{\rm s}/\lambda_{\rm J}$. The resulting escape fraction is closer to the true solution than is the case for MH-Sob, but nevertheless deviates by a factor of a few at both ends of the profile. This is due to the variation of the gas properties on scales smaller than the velocity varies, which violates the Sobolev condition.

The difference between MH-Sob-M0 and MH-Sob shows that large velocity gradients induced by the transonic turbulence throughout the cloud reduce the Sobolev length by a factor of a few compared to the Jeans length, resulting in a similar increase in the escape fraction. This is because the Sobolev method assumes that the local velocity gradient is coherent throughout the entire cloud, while in reality the velocity can vary significantly on scales smaller than the Jeans length, and has a mean that is approximately equal to the net infall velocity. If the Sobolev method is used, the turbulence thus has a strong effect on the escape fraction, while a comparison of MH-Ray-M2 and MH-Ray in the top panel of Fig.~\ref{fig_mh_fesc} and the results of Fig.~\ref{fig_mh_freq} shows that the turbulence does not have a substantial effect in the case of MH-Ray. The discrepancy in the escape fraction between MH-Sob and MH-Ray at the highest densities can thus exceed more than an order of magnitude. The fitting function of \citet{ra04}, on the other hand, agrees relatively well with the ray tracing during the initial collapse phase. In general, however, this will not be the case since the escape fraction depends on many factors, such as the density, temperature, velocity, the chemical and thermal rate equations, and the further evolution of the cloud.

\subsection{Comparison to previous work}

We here compare the results of our simulations to previous work. \citet{yoshida06b} investigated the collapse of primordial gas clouds with smoothed particle hydrodynamics (SPH) simulations, using the Sobolev method to estimate the photon escape fraction. They found an escape fraction of $\simeq 0.02$ at $n_{\rm H}=10^{14}\,{\rm cm}^{-3}$, which agrees reasonably well with MH-Ray at this density, where $f_{\rm esc}\simeq 0.01$. However, this does not agree with MH-Sob, where the escape fraction is an order of magnitude higher. Since the radial velocities are comparable in both studies, this difference is likely caused by the turbulence of the gas. This turbulence was not resolved in the simulations of \citet{yoshida06b}, due to inherent limitations of the hydrodynamic solver employed \citep{bs12}. \citet{turk11} compared the escape fraction obtained with the Sobolev method in an SPH simulation to that described by the fitting function of \citet{ra04}. They found that the resulting escape fractions differed by about a factor of $2$ in the range $10^{10}\la n_{\rm H}\la 10^{15}\,{\rm cm}^{-3}$. A somewhat more detailed study was carried out by \citet{hy13}. They found gas clouds that were substantially rotationally supported and had a smaller radial velocity gradient than in the one-dimensional calculations of \citet{on98} and \citet{ripamonti02}. As a result, they obtained escape fractions that were systematically smaller than those described by the fitting function. However, similar to \citet{yoshida06b}, both studies did not resolve the turbulence of the gas very well.

Despite the very different escape fractions obtained with the various methods, the thermal evolution of the clouds does not differ by much. This is due to the strong temperature dependence of the H$_2$ line cooling rate, which scales approximately as $\Lambda_{\rm LTE}\propto T^4$. Even for an order of magnitude difference in the escape fraction, the temperature thus varies by less than a factor of $2$. For example, \citet{gsb13} employed the Sobolev method using a moving-mesh approach and found an average temperature of $\simeq 1800\,{\rm K}$ at $n_{\rm H}\simeq 10^{15}\,{\rm cm}^{-3}$, while \citet{tao09} used the fitting function and found a central temperature of $\simeq 2500\,{\rm K}$. Other studies found values between these two extremes \citep{clark11b, greif11a, greif12, turk11, turk12, hy13}. Here, we find that the central temperatures obtained with the various methods are in the range $2200$--$2300\,{\rm K}$, despite the highly discrepant escape fractions. Next to the temperature, the H$_2$ abundance varies as well. For example, in \citet{gsb13} the gas remains fully molecular at $n_{\rm H}=10^{15}\,{\rm cm}^{-3}$, while in \citet{tao09} the H$_2$ abundance drops to $y_{\rm H_2}\simeq 0.2$. In the present study, the H$_2$ abundance only varies between $\simeq 0.3$ and $0.4$. We do find, however, a large difference in the dispersion of the temperature at a density of $n_{\rm H}\simeq 10^{12}\,{\rm cm}^{-3}$. In the most extreme case, \citet{gsb13} found that the temperature can vary by a factor of $\simeq 5$ and that some parcels of gas become gravitationally unstable. In the present study, the temperature varies by less than a factor of $2$, demonstrating that the thermal instability found in \citet{gsb13} is caused by the overestimate of the escape fraction by the Sobolev method. Since the turbulence was well resolved in \citet{gsb13}, the temperature dispersion is even larger than in studies that employed SPH simulations \citep[e.g.][]{yoshida06b, clark11b}.

We note that another potential source of discrepancy is the rate used for three-body H$_2$ formation and the inverse process, collisional dissociation. Nearly all previous studies used the rates introduced in \cite{glover08}, which are intermediate in terms of the large uncertainty discussed in \citet{turk11}. In the present study, we use the revised three-body formation rate obtained by the quantum-mechanical calculations of \citet{forrey13}, which is about two times lower than the rate of \citet{glover08} at $1000\,{\rm K}$. Since the difference is significantly smaller than in the case of the escape fractions, it likely does not have a substantial effect on the thermal evolution of the gas.

\section{Summary and Conclusions} \label{sec_sum}

We have performed the first three-dimensional simulations of primordial star formation that self-consistently model the multifrequency radiative transfer of H$_2$ line emission. The simulations employ a new equilibrium/non-equilibrium primordial chemistry solver next to a new multiline, multifrequency ray-tracing scheme that is capable of adaptively refining rays based on the {\sc healpix} algorithm \citep{gorski05}. The latter can be used to solve the static radiative transfer equation for point sources as well as diffuse emission. Both schemes have been implemented in the simulation code {\sc arepo}. The chemistry solver is optimized for collapse simulations and is significantly faster than the solver used in \citet{greif12}. The ray-tracing scheme is capable of walking about one million cells per second, and can be parallelized using a hybrid distributed/shared memory scheme. The calculation of the optical depth for multiple lines and/or frequency bins uses tabulated opacities and exploits the Intel AVX instruction set, which boosts the performance to a level where the main bottleneck is the communication of the ray data among {\sc MPI} tasks. For isolated point sources, the parameters that govern the accuracy of the scheme are the initial {\sc healpix} level, the average number of rays per cell (for adaptive splitting), the minimum fractional energy of a ray before it is terminated, the number of lines, and the number of frequency bins. In the case of diffuse emission, the angular resolution typically remains constant, while an additional parameter specifies the fraction of all cells that are sources. The reliability of both schemes is demonstrated with a series of idealized test calculations.

The ray-tracing scheme is used to compute the radiative transfer of H$_2$ line emission in an ab initio simulation of primordial star formation. We achieve an accuracy of $5$ per cent in the radiative heating rate by using $\simeq 10^{14}$ opacity calculations per time step, amounting to a total wall-clock time of $1$--$2$ months on $1024$ state-of-the-art computing cores. In agreement with previous studies, we find that the gas becomes optically thick to H$_2$ line emission at densities $n_{\rm H}\ga 10^{10}\,{\rm cm}^{-3}$, and the line cooling rate is surpassed by collision-induced emission at $n_{\rm H}\ga 10^{15}\,{\rm cm}^{-3}$. Within this range, the spherically averaged escape fraction decreases from unity to $\simeq 2\times 10^{-3}$, with a power-law slope of $\simeq -0.6$. This agrees relatively well with the fitting function of \citet{ra04}, which is based on one-dimensional radiative transfer calculations \citep{on98, ripamonti02}. During the initial collapse phase, the assumption of spherical symmetry appears to give relatively accurate results for the purpose of computing the H$_2$ line transfer. However, since the escape fraction depends on many factors, such as the density, velocity, temperature, and the chemical and thermal rate equations, it is generally not advisable to use a fitting function. The escape fraction method also does not capture the diffusion of the radiation, which suppresses density fluctuations as the gas evolves into the optically thick regime.

By systematically increasing the physical detail of the radiative transfer, we have found that using multiple lines and frequency bins is essential. The lower cross-sections of the more sparsely populated lines can boost the amount of energy that can escape by many orders of magnitude. A similar effect becomes important if frequency-dependent emission and absorption are accounted for: the lower cross-sections in the wings of the lines allow significantly more energy to escape than in the grey case. Finally, Doppler shifts due to relative velocities along the rays increase the escape fraction by about a factor of $2$. This effect is relatively small in comparison, since the infall velocity fluctuates by less than the sound speed, and results in a frequency shift of only $\simeq \Delta\nu_{\rm D}/2$.

We have also compared our results to the escape fraction obtained with the Sobolev method. For low optical depths, the Sobolev method somewhat underestimates the escape fraction, while for high optical depths the escape fraction is overestimated by more than an order of magnitude. This discrepancy arises because the Sobolev method is only accurate if the scales on which the properties of the gas change are much larger than the Sobolev length. This is not the case in the self-gravitating gas clouds that form in minihaloes, since the infall velocity typically varies by less than the sound speed. The discrepancy becomes even larger if the turbulence is well resolved. In this case, the local velocity gradient can be much larger than $c_{\rm s}/\lambda_{\rm J}$, resulting in a further increase in the escape fraction, despite the fact that the turbulent velocities nearly cancel each other within a Jeans length, and thus have almost no effect on the escape fraction. Previous studies found better agreement between the Sobolev method and the fitting function of \citet{ra04}, since limitations of the hydrodynamic solver employed prevented the turbulence from being resolved \citep{yoshida06b, clark11b, turk11, hy13}. As a result, the velocity gradient was dominated by the radial velocity gradient, resulting in a Sobolev length that was significantly larger than the one found here.

For the above reasons, simulations that used the Sobolev method and resolved the turbulence in the gas greatly overestimated the escape fraction \citep{greif12, gsb13}. In particular, the cooling instability found in \citet{gsb13} is largely an artefact of the Sobolev method. However, due to the strong dependence of the H$_2$ line cooling rate on the temperature, the overall thermal evolution of the cloud is much less affected. For example, in \citet{gsb13} the temperature at $n_{\rm H}\simeq 10^{15}$ is $\simeq 1800\,{\rm K}$, while in \citet{tao09} the temperature is $\simeq 2500\,{\rm K}$. The H$_2$ abundance shows a similarly mild variation. In \citet{gsb13}, the gas remains fully molecular at a density of $\simeq 10^{15}$, while in \citet{tao09} the H$_2$ has begun to dissociate, with $y_{\rm H_2}\simeq 0.2$. \citet{turk12} suggested that the reduced H$_2$ fraction may reduce the ability of the cloud to fragment. However, since the gas becomes rotationally supported in a Keplerian disc following the initial collapse, the resulting asymmetry may allow the cooling radiation to escape more easily than previous studies predicted \citep{clark11a, greif11a, greif12}. A definitive answer must await detailed radiation hydrodynamics simulations that evolve the collapse well beyond the formation of the first protostar.

\section*{Acknowledgements}
THG is indebted to Volker Springel for access to the simulation code {\sc arepo}. THG would like to thank Lars Hernquist, Laura Sales, Mark Vogelsberger, Volker Bromm, and Volker Springel for stimulating discussions and feedback. The simulations were carried out at the Texas Advanced Computing Center (TACC) under XSEDE allocation AST130020.

\end{document}